%% file: Main.tex
\documentclass[aps,pre,twocolumn,superscriptaddress,showpacs,longbibliography]{revtex4-1}
\usepackage{amsmath}
\usepackage{amssymb}
\usepackage{bm}
\usepackage{braket}
\usepackage{multirow}
\usepackage{comment}
\usepackage[normalem]{ulem}
\usepackage{here} % for position of figures
\usepackage{graphicx}
\usepackage{color}
\usepackage[colorlinks=true, linkcolor=magenta,citecolor=blue,urlcolor=magenta]{hyperref}
\DeclareGraphicsExtensions{.pdf}	%%% Extensions for pdflatex
\newcommand{\Tr}[1]{\text{Tr}[#1]}

\newcommand{\Ex}[1]{\langle #1 \rangle}

\def\eps{\epsilon}
\def\veps{\varepsilon}

\def\F{\mathrm F}

\def\R{\mathrm R}
\def\S{\mathrm S}
\def\B{\mathrm B}

\def\e{\mathrm e}
\def\c{\mathrm c}
\def\s{\mathrm s}
\def\x{\mathrm x}
\def\y{\mathrm y}
\def\z{\mathrm z}
\def\d{\mathrm d}
\def\ph{\mathrm{ph}}
\def\tot{\mathrm{tot}}

\def\ds{\delta_{\s}}

\def\Tr{\mathrm{Tr}}
\def\Real{\mathrm{Re}}

\def\free{\mathrm{free}}
\def\ii{\mathrm i}
\def\dd{\mathrm d}

\def\oE{\hat{\bm{E}}}
\def\uE{\breve{\bm{E}}}

\def\uI{\breve{I}}
\def\mI{\mathcal{I}}

\def\oH{\hat{H}}
\def\oHS{\hat{H}_{\mathrm S}}
\def\oHB{\hat{H}_{\mathrm B}}
\def\oHSB{\hat{H}_{\mathrm{SB}}}
\def\vHSB{\check{H}_{\mathrm{SB}}}

\def\osgm{\hat{\sigma}}
\def\vsgm{\check{\sigma}}
\def\usgm{\breve{\sigma}}

\def\oac{\hat{a}_{\c}}
\def\oacd{\hat{a}_{\c}^{\dagger}}
\def\vac{\check{a}_{\c}}

\def\uac{\breve{a}_{\c}}
\def\uacd{\breve{a}_{\c}^{\dagger}}

\def\ob{\hat{b}}
\def\obd{\hat{b}^{\dagger}}
\def\vb{\check{b}}

\def\ub{\breve{b}}

\def\orho{\hat{\rho}}
\def\vrho{\check{\rho}}

\def\oV{\hat{V}}
\def\vV{\check{V}}
\def\oO{\hat{O}}
\def\vO{\check{O}}
\def\uO{\breve{O}}

\def\vD{\check{\mathcal{D}}}

\def\meV{\text{meV}}
\def\ueV{\mu\text{eV}}

\begin{document}
%===========================
%   Title, Author, Affiliation
%===========================
\title{Theory of Fano effect in cavity quantum electrodynamics}

\author{Makoto Yamaguchi}
\altaffiliation{E-mail: makoto.yamaguchi@tokai.ac.jp}
\affiliation{Department of Physics, Tokai University, 4-1-1 Kitakaname, Hiratsuka, Kanagawa 259-1292, Japan}
\author{Alexey Lyasota}
\affiliation{Laboratory of Physics of Nanostructures, Institute of Physics, Ecole Polytechnique F\'{e}d\'{e}rale de Lausanne (EPFL), Lausanne, Switzerland}
\affiliation{Centre of Excellence for Quantum Computation and Communication Technology, School of Physics, University of New South Wales, Sydney, New South Wales 2052, Australia}
\author{Tatsuro Yuge}
\affiliation{Department of Physics, Shizuoka University, Shizuoka 422-8529, Japan}

\date{\today}

%======================================================
%   Abstract (no more than 600 characters)
%======================================================
\begin{abstract}
We propose a Markovian quantum master equation that can describe the Fano effect directly, by assuming a standard cavity quantum electrodynamics system.
The framework allows us to generalize the Fano formula, applicable over the weak and strong coupling regimes with pure dephasing.
A formulation of its emission spectrum is also given in a consistent manner.
We then find that the interference responsible for the Fano effect is robust against pure dephasing.
This is counterintuitive because the impact of interference is, in general, severely reduced by decoherence processes.
Our approach thus provides a basis for theoretical treatments of the Fano effect and new insights into the quantum interference in open quantum systems.
\end{abstract}
\maketitle

%======================================================
%   Introduction
%======================================================
\section{Introduction}\label{sec:Intro}
\begin{figure}
\centering
\includegraphics[width=.50\textwidth]{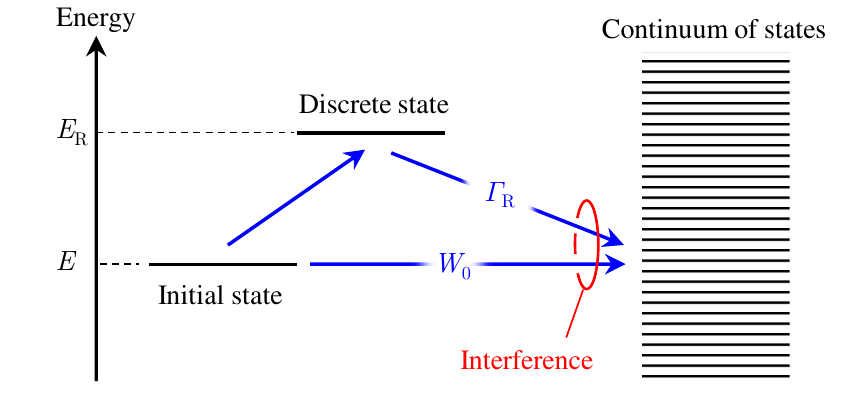}
\caption{
    Schematic illustration of the interference for the Fano effect.
    The initial state has two dissipation channels into the continuum of states, which can interfere with each other.
    $E$ is the energy of the initial state, while $E_{\R}$ and $\Gamma_{\R}$ are the resonance energy and width of the discrete state, respectively.
    $W_{0}$ is the transition rate only due to the direct channel.
}
\label{fig:Fano}
\end{figure}

The Fano effect provides one of the key insights for the study of resonance physics~\cite{Fano61,Joe06,Miroshnichenko10,Limonov17}.
The mechanism of this effect requires only a few ingredients, the generality of its concept is significant, and the predictions obtained are simple but critical to understand a transition rate from an arbitrary initial state that has two dissipation channels with interference; one is a direct channel to a continuum band of states and the other is an indirect channel via a discrete state into the same continuum (Fig.~\ref{fig:Fano}).
As a result, the transition rate from the initial state given by the Fano formula,
%----------
% Equation
%----------
\begin{align}
    W_{\F}=W_{0}\frac{(q+\eps)^2}{1+\eps^2},
    \label{eq:Fano}
\end{align}
%----------
has a wide range of applications, such as atomic physics~\cite{Fano86}, Raman scattering~\cite{Cerdeira73,Magidson02}, lasing without inversion~\cite{Harris89,Imamoglu89}, gain spectra in semiconductors~\cite{Kamide12}, and photonic systems~\cite{Fan,Fan03}, especially when focused on the scattering (gain or absorption) problems for externally introduced field.
The asymmetric resonance profile of $W_{\F}$ as a function of the reduced energy $\eps \equiv 2(E-E_{\R})/\Gamma_{\R}$ is now known as the characteristic sign of the Fano effect, where the degree of the asymmetry is determined by the Fano parameter $q$; see also Fig.~\ref{fig:Fano} for the definitions of relevant variables.

In general, however, the Fano effect is inherently not limited to such scattering (gain or absorption) problems.
Recent studies have pointed out that the interference effect, indeed, plays important roles even in the simplest situations of cavity quantum electrodynamics (QED)~\cite{Yamaguchi08,Barclay09,Ota15}, where an initially excited two-level system (TLS) spontaneously emits a photon into a continuum of radiation modes, directly and indirectly via a single mode cavity.
Nevertheless, there are few theoretical studies to tackle the Fano effect, based on the modern theories of open quantum systems~\cite{Scully97,Carmichael99,Breuer02,Scully97,Carmichael89,Cui06,Laussy08,Yamaguchi12}; the interference between the dissipative channels is outside the scope of these studies.
As a result, for example, it is difficult to discuss the influence of the strong coupling and/or the pure dephasing on the Fano effect.
Furthermore, the emission spectra can also be modified as a result of the interference~\cite{Barclay09,Ota15,Madsen13}.
In this context, a more flexible and systematic theory has increasing importance for understanding the Fano effect in open quantum systems~\cite{Denning19,Cernotik19,Franke19}.

In this paper, we propose a Markovian quantum master equation (QME) that provides a simple approach for the description of the Fano effect, by assuming an ideal cavity QED system.
We find that the interference effect can be directly implemented into a Liouville superoperator.
As a result, an analytical expression of the transition rate, $W$, is obtained over the weak and strong coupling regimes with pure dephasing, as a generalization of the Fano formula [Eq.~\eqref{eq:Fano}].
Furthermore, a formulation of its emission spectrum is also given in a consistent manner with our treatment of the QME.
Then, despite the smeared transition rate by the pure dephasing, we find that the interference effect is itself rather insensitive to the pure dephasing, based on the calculated spectra.
As a result, the destructive interference can eliminate the emission line at the TLS transition energy with the help of pure dephasing.
This is in contrast to a naive intuition that the impact of interference is severely reduced by decoherence processes in general.
The underlying physics is elucidated by studying the fundamental mechanism of the Fano effect from the viewpoint of spectra.
Our scheme thus proposes a Markovian QME approach for the Fano effect, achieves a generalization of the Fano formula,  and gives new insights into the quantum interference of the Fano effect.
In consequence, our results can provide a basis for theoretical treatments of the Fano effect in open quantum systems.

The rest of the paper is organized as follows.
In Sec.~\ref{sec:Setup}, we describe our setup of the cavity QED system, where the TLS is coupled with the single mode cavity.
At this stage, several assumptions are introduced to discuss the Fano effect.
In Sec.~\ref{sec:QME}, we explain the Liouville superoperator that describes the interference effect in the QME.
We then derive the transition rate, $W$, by assuming that the TLS is initially excited.
In Sec.~\ref{sec:Spectra}, we formulate the emission spectra in a consistent way with our treatment of QME.
Based on numerical calculations, then, we discuss the Fano effect on the emission spectra.
In Sec.~\ref{sec:Conclusion}, we summarize our results.
Throughout the paper, we set $\hbar = 1$ for simplicity.

%----------
% Figure
%----------
%----------

%======================================================
%   Formulation
%======================================================
\section{Setup}\label{sec:Setup}
Our theoretical stage is a standard cavity QED system positioned at the origin of spatial coordinates, where a TLS with transition energy $\omega_{21}$ can interact with a single mode cavity with resonant energy $\omega_{\c}$ by a coupling constant $g$.
We assume that the TLS and the cavity mode have decay rates of $\gamma$ and $\kappa$, respectively, as a result of the interaction with a continuum of radiation modes in the environment.

In order to formulate the QME, we describe the total Hamiltonian $\oH$ as
%----------
% Equation
%----------
\begin{align}
    \oH = \oHS + \oHB + \oHSB,
    \label{eq:Htot}
\end{align}
%----------
in the Schr\"odinger picture, where $\oHS$ is the Hamiltonian of the system, $\oHB$ is the Hamiltonian of the baths, and $\oHSB$ is the interaction Hamiltonian between the system and the baths.
We consider the TLS and the cavity mode as the system, and the continuum of radiation modes as one of the baths.
Hence, the system Hamiltonian, $\oHS$, is given by
%----------
% Equation
%----------
\begin{align}
    \oHS=\frac{1}{2}\omega_{21}\osgm_{\z} + \omega_{\c}\oacd\oac + (g\osgm_{+}\oac + g^{*}\oacd\osgm_{-}),
    \label{eq:HS}
\end{align}
%----------
where $\oacd$ and $\oac$ are the bosonic creation and annihilation operators of the cavity photons, $\osgm_{+}$ and $\osgm_{-}$ are the raising and lowering operators of the TLS, and $\osgm_{i}$ ($i = \x, \y, \z$) is the Pauli operator of the TLS~\cite{Scully97}.
Here, $g$ is described by a complex number, $g = |g|e^{\ii\phi}$ with $\phi = \pi/2$.
In a similar manner, the Hamiltonian of the baths, $\oHB$, is described as  
%----------
% Equation
%----------
\begin{align}
\oHB=\sum_{\ell} \omega_{\ell} \obd_{\ell}\ob_{\ell} + \cdots,
\label{eq:HB}
\end{align}
%----------
where $\obd_{\ell}$ and $\ob_{\ell}$ are the bosonic creation and annihilation operators of a radiation mode $\ell$ with its energy $\omega_{\ell}$.
Here, $\ell$ denotes the wavevector and the polarization, $(\bm{k}, \lambda)$, in the continuum.
Contributions of other baths, responsible for the pure dephasing, are not shown in Eq.~\eqref{eq:HB} to avoid digressing from the main subject.
For the interaction Hamiltonian, $\oHSB$, we have
%----------
% Equation
%----------
\begin{align}
    \oHSB=\oHSB^{(1)}+\oHSB^{(2)}+\cdots,
    \label{eq:HSB}
\end{align}
%----------
with
%----------
% Equation
%----------
\begin{align}
    \oHSB^{(1)}=\sum_{\ell} (\xi_{\ell}\osgm_{+}\ob_{\ell} + \xi^{*}_{\ell}\obd_{\ell}\osgm_{-}),
    \label{eq:HSB1}\\
    \oHSB^{(2)}=\sum_{\ell} (\zeta_{\ell}\oacd\ob_{\ell} + \zeta^{*}_{\ell}\obd_{\ell}\oac),
    \label{eq:HSB2}
\end{align}
%----------
where $\xi_{\ell}$ ($\zeta_{\ell}$) is the coupling constant between the TLS (the cavity mode) and the $\ell$-th radiation mode.
We describe $\xi_{\ell} = |\xi_{\ell}|e^{\ii\theta_{21}}$ and $\zeta_{\ell} = |\zeta_{\ell}|e^{\ii\theta_{\c}}$ as complex numbers, where $\theta_{21} = \pi/2$ in the same manner as $g$ and $\theta_c$ is unknown in general.
Again, other interaction Hamiltonians are not shown in Eq.~\eqref{eq:HSB} for brevity.

In Sec.~\ref{sec:QME}, we will discuss the QME, based on these Hamiltonians.
Before preceding further, however, we make the following three assumptions:
\begin{enumerate}
    \item The absolute value of the detuning, $\omega_{\c,21} \equiv \omega_{\c} - \omega_{21}$, under consideration is much smaller than $\omega_{21}$ and $\omega_{\c}$.
    \item We assume that the coupling constants, $\xi_{\ell}$ and $\zeta_{\ell}$, can be simplified as a function of $\omega_{\ell}$, i.e., $\xi_{\ell} \simeq \bar{\xi}(\omega_{\ell})$ and $\zeta_{\ell} \simeq \bar{\zeta}(\omega_{\ell})$.
    \item $\bar{\xi}(\omega)$, $\bar{\zeta}(\omega)$, and $D(\omega) \equiv \sum_{\ell} \delta(\omega - \omega_{\ell})$ do not depend strongly on the energy $\omega$.
    Here, $D(\omega)$ is the density of states of the continuum.
\end{enumerate}
As a result, the decay rates of the TLS and the cavity mode are given by 
%----------
% Equation
%----------
\begin{align}
    \gamma=2\pi|\bar{\xi}(\omega_{21})|^2 D(\omega_{21}), \quad
    \kappa=2\pi|\bar{\zeta}(\omega_{\c})|^2 D(\omega_{\c}),
    \label{eq:gamma and kappa}
\end{align}
%----------
respectively.
In fact, similar assumptions are implicitly used in Fano's original work~\cite{Fano61}.
However, we note that the second assumption is drastic because, in general, $\xi_{\ell}$ and $\zeta_{\ell}$ depend on the direction of the wavevector and the polarization of the $\ell$-th mode as well as its energy.
Nevertheless, we employ this assumption to extract the essential features.
$\bar{\xi}(\omega)$ and $\bar{\zeta}(\omega)$ correspond to the coupling constants averaged over all directions.
In addition, based on the first and third assumptions, we ignore the dependence of $\gamma$ and $\kappa$ on $\omega_{21}$ and $\omega_{\c}$, in the following.

\section{The QME approach to the Fano effect}\label{sec:QME}
\subsection{A formulation of the QME}\label{subsec:QME}
Now, we formulate the QME.
For this purpose, we introduce the interaction picture with respect to $\oH_{0} \equiv \frac{1}{2}\omega_{21}\osgm_{\z} + \omega_{\c}\oacd\oac + \oH_{\B}$,
%----------
% Equation
%----------
\begin{align*}
    \vO(t) \equiv e^{\ii\oH_{0}t} \oO e^{-\ii\oH_{0}t},
\end{align*}
%----------
where $\oO$ is an arbitrary operator.
Under the Born-Markov approximation, then, the time evolution of the reduced density operator of the system, $\vrho_{\S} \equiv \Tr_{\B}[\vrho]$, is given by
%----------
% Equation
%----------
\begin{align}
    \frac{\dd}{\dd t}\vrho_{\S}(t) &= -\ii[\vV_{\S}(t), \vrho_{\S}(t)] \nonumber\\ 
    &-\int^{\infty}_{0}\dd\tau \Tr_{\B} [\vHSB(t) ,[\vHSB(t-\tau), \vrho_{\S}(t) \otimes \orho_{\B}]],
    \label{eq:BM-QME}
\end{align}
%----------
where $\vV_{\S}(t)$ is the interaction picture of $\oV_{\S} \equiv g\osgm_{+}\oac + g^{*}\oacd\osgm_{-}$, $\orho_{\B}$ is the density operator of the baths, and we have used $\Tr_{\B}[\oHSB \orho_{\B}] = 0$ by assuming the continuum of radiation modes is in vacuum.
Here, we note that the reference Hamiltonian of the interaction picture is $\oH_{0}$, instead of $\oH_{\S}+\oH_{\B}$, and therefore, Eq.~\eqref{eq:BM-QME} is slightly different from the standard approach~\cite{Breuer02}; see also Appendix~\ref{app:BM-QME}.
As we see below, this enables us to directly use the system operators of $\oHSB$, such as $\oac$, $\oacd$, and $\osgm_{\pm}$, in the final form of dissipators because $\vHSB(t')$ can be described by
$\vac(t')=\oac e^{-\ii\omega_{\c}t'}$,
$\vsgm_{-}(t')=\osgm_{-}e^{-\ii\omega_{21}t'}$, 
$\vb_{\ell}(t')=\ob_{\ell} e^{-\ii\omega_{\ell}t'}$, 
and their Hermitian conjugates.

To derive the dissipators, we substitute Eq.~\eqref{eq:HSB} into the double commutator in Eq.~\eqref{eq:BM-QME};
%----------
% Equation
%----------
\begin{align}
    &[\vHSB(t) ,[\vHSB(t-\tau), \bullet]] \nonumber\\
    &\quad = [\vHSB^{(1)}(t) ,[\vHSB^{(1)}(t-\tau), \bullet]]
     + [\vHSB^{(2)}(t) ,[\vHSB^{(2)}(t-\tau), \bullet]]\nonumber\\
    &\quad + [\vHSB^{(1)}(t) ,[\vHSB^{(2)}(t-\tau), \bullet]]
     + [\vHSB^{(2)}(t) ,[\vHSB^{(1)}(t-\tau), \bullet]]\nonumber\\
    &\quad + \cdots.
    \label{eq:DoubleCommutation}
\end{align}
%----------
This equation means that the dissipators will be generated not only by the direct terms (the first and second terms) but also the cross terms (the third and forth terms) of the individual interaction Hamiltonians [Eqs.~\eqref{eq:HSB1} and \eqref{eq:HSB2}].
In fact, in Eq.~\eqref{eq:BM-QME}, the direct terms give the well-known dissipators that describe the dissipative effect,
%----------
% Equation
%----------
\begin{align}
    &\vD_{21}\vrho_{\S} = \frac{\gamma}{2}(2\osgm_{-}\vrho_{\S}\osgm_{+} - \osgm_{+}\osgm_{-}\vrho_{\S} - \vrho_{\S}\osgm_{+}\osgm_{-} ),
    \label{eq:D21}\\
    &\vD_{\c}\vrho_{\S} = \frac{\kappa}{2}(2\oac\vrho_{\S}\oacd - \oacd\oac\vrho_{\S} - \vrho_{\S}\oacd\oac ),
    \label{eq:Dc}
\end{align}
%----------
by following the standard procedures to obtain the QME~\cite{Scully97,Carmichael99}.
For simplicity, we neglect the terms corresponding to the Lamb shift in our discussion.
In contrast, the cross terms yield
%----------
% Equation
%----------
\begin{align}
    \vD_{\F}\vrho_{\S} &= \frac{\gamma_{\F}}{2}e^{-\ii\omega_{\c,21}t}(2\oac\vrho_{\S}\osgm_{+} - \osgm_{+}\oac\vrho_{\S} - \vrho_{\S}\osgm_{+}\oac ) \nonumber\\
    &+\frac{\gamma^{*}_{\F}}{2}e^{\ii\omega_{\c,21}t} (2\osgm_{-}\vrho_{\S}\oacd - \oacd\osgm_{-}\vrho_{\S} - \vrho_{\S}\oacd\osgm_{-} ),
    \label{eq:FanoDissipator}
\end{align}
%----------
where $\gamma_{\F}$ is a complex number given by 
%----------
% Equation
%----------
\begin{align}
    \gamma_{\F} = e^{\ii(\theta_{21}-\theta_{\c})}\sqrt{\eta\gamma\kappa}.
    \label{eq:gamma_F}
\end{align}
%----------
In the derivation, we have used the assumptions described in Sec.~\ref{sec:Setup} and introduced a phenomenological parameter $\eta$ ($0 \le \eta \le 1$) that describes the degree of overlap between the spatial radiation patterns of the TLS and the cavity mode; see also Appendx~\ref{app:FanoDissipator}. 
Here, $\eta=1$ for the identical radiation patterns, while $\eta=0$ for the orthogonal ones.
The superoperator $\vD_{\F}$ is usually neglected by implicitly assuming the orthogonal radiation patterns ($\eta=0$).
Here, in contrast to $\gamma$ and $\kappa$, we note that $\gamma_{\F}$ is a complex number, the phase of which is determined by $\theta_{21}$ and $\theta_{\c}$.
Therefore, it is important to treat the coupling constants as complex numbers in the original interaction Hamiltonians [Eqs.~\eqref{eq:HSB1} and \eqref{eq:HSB2}], while such a treatment is not required for the description of $\vD_{21}$ [Eq.~\eqref{eq:D21}] and $\vD_{\c}$ [Eq.~\eqref{eq:Dc}].
To our knowledge, this is the first time such a dissipator is presented despite its simple form.
We note that extension of our formulation to a finite temperature case is straightforward although the continuum was assumed in vacuum.

In addition to the dissipators shown above, in the following, we also use a dissipator that describes the pure dephasing effect,
%----------
% Equation
%----------
\begin{align}
    \vD_{\ph}\vrho_{\S} = \frac{\gamma_{\ph}}{2}(\osgm_{\z}\vrho_{\S}\osgm_{\z}-\vrho_{\S}).
    \label{eq:Dephasing}
\end{align}
%----------
where $\gamma_{\ph}$ is the pure dephasing rate of the TLS~\cite{Carmichael99}.
As a result, the QME in our study is finally given by 
%----------
% Equation
%----------
\begin{align}
    \frac{\dd}{\dd t}\vrho_{\S} &= -\ii[\vV_{\S}, \vrho_{\S}] 
    +(\vD_{21}+\vD_{\c}+\vD_{\F}+\vD_{\ph})\vrho_{\S},
    \label{eq:QME}
\end{align}
%----------
in the interaction picture.
At this stage, however, it is difficult to understand the effect of the dissipator, $\vD_{\F}$, [Eq.~\eqref{eq:FanoDissipator}].
In the next section, therefore, we study the transition rate, based on the QME described here.

\subsection{The transition rate}\label{subsec:Transition rate}
To clarify the effect of the dissipator, $\vD_{\F}$, we now discuss the transition rate of the initially excited TLS.
For this purpose, we start with the time evolution of relevant expectation values, $\Ex{\oO}_{t} \equiv \Tr[\oO\orho(t)]$.
After transforming back into the Schr\"odinger picture, the QME yields the equations of motion for the population of the excited state, $n_{\e}(t) \equiv \braket{\osgm_{+}\osgm_{-}}_{t}$, the photons inside the cavity, $n_{\c}(t) \equiv \braket{\oacd\oac}_{t}$, and the polarization, $p(t) \equiv \braket{\osgm_{+}\oac}_{t}$;
%----------
% Equation
%----------
\begin{align}
    &\dot{n}_{\c} = 2\Real(\ii g_{+}p) - \kappa n_{\c},
    \label{eq:n_c}\\
    &\dot{n}_{\e} = 2\Real(-\ii g_{-}p) - \gamma n_{\e},
    \label{eq:n_e}\\
    &\dot{p} = -\ii g^{*}_{+} n_{\e} + \ii g^{*}_{-} n_{\c} - ( \ii\omega_{\c,21} + \Gamma_{\tot} )p,
    \label{eq:p}
\end{align}
%----------
where $g_{\pm}$ and $\Gamma_{\tot}$ are respectively defined by
%----------
% Equation
%----------
\begin{align}
    &g_{\pm} \equiv g \pm \ii \frac{\gamma_{\F}}{2},
    \label{eq:g+-}
\end{align}
%----------
and
%----------
% Equation
%----------
\begin{align}
    &\Gamma_{\tot} \equiv \frac{\gamma+\kappa}{2} + \gamma_{\ph}.
    \label{eq:Gamma_tot}
\end{align}
%----------
Here, an approximation $\braket{\osgm_{\z}\oacd\oac}_{t} \simeq - \braket{\oacd\oac}_{t}$ has been introduced in Eq.~\eqref{eq:p} because, in our situation, the TLS is always in the ground state when a photon is inside the cavity~\cite{Scully97,Yamaguchi12}.
In these equations, we can notice that the TLS can interact with the cavity mode through the continuum of states because $\gamma_{\F}$ plays a similar role to the coupling constant $g$.
However, such an interaction cannot reduce to a simple renormalization of $g$.
We need $g_{+}$ and $g_{-}$ to describe the different ways of coupling between $p$ and $n_{\c}$, and that between $p$ and $n_{\e}$, as seen in Eqs.~\eqref{eq:n_c}--\eqref{eq:p}.

Here, it is not easy to obtain the analytical solutions for Eqs.~\eqref{eq:n_c}--\eqref{eq:p} although these equations can give the time evolution over the weak and strong coupling regimes.
For our purpose to obtain the transition rate, $W$, however, there is no need to exactly solve the problem.
In the strong coupling regime, in particular, the transition rate can be measured from the decay of the envelope of the Rabi oscillations~\cite{Yamaguchi12}.
In this context, to extract the transition rate, we further introduce an approximation, $\dot{p} \simeq 0$, in Eq.~\eqref{eq:p};
%----------
% Equation
%----------
\begin{align}
    p \simeq \frac{-\ii g^{*}_{+} n_{\e} + \ii g^{*}_{-} n_{\c}}{\ii \omega_{\c, 21} + \Gamma_{\tot}},
    \label{eq:Adiabatic}
\end{align}
%----------
which averages out the Rabi oscillations if the system is in the strong coupling regime~\cite{Note1}.
The coarse-grained time evolution is then described by  
%----------
% Equation
%----------
\begin{align}
    &\dot{n}_{\c} = -(R_{+,-}+\kappa)n_{\c} + R_{+,+}n_{\e},
    \label{eq:n_c2}\\
    &\dot{n}_{\e} = -(R_{-,+}+\gamma)n_{\e} + R_{-,-}n_{\c},
    \label{eq:n_e2}
\end{align}
%----------
where
%----------
% Equation
%----------
\begin{align}
    R_{\alpha, \beta} \equiv \Real \left( \frac{2g_{\alpha}g^{*}_{\beta}}{\ii\omega_{\c,21}+\Gamma_{\tot}} \right).
\end{align}
%----------
As a result, Eqs.~\eqref{eq:n_c2} and \eqref{eq:n_e2} with the initial conditions, $n_{\c}(0)=0$ and $n_{\e}(0)=1$, can yield the analytic solutions,
%----------
% Equation
%----------
\begin{align}
    &n_{\e}(t)=\frac{\Lambda_{+}+R_{+,-}+\kappa}{\Lambda_{+}-\Lambda{-}}e^{\Lambda_{+}t} - \frac{\Lambda_{-}+R_{+,-}+\kappa}{\Lambda_{+}-\Lambda{-}}e^{\Lambda_{-}t},
    \label{eq:ne(t)}\\
    &n_{\c}(t)=\frac{R_{+,+}}{\Lambda_{+}-\Lambda{-}}(e^{\Lambda_{+}t} - e^{\Lambda_{-}t}),
    \label{eq:nc(t)}
\end{align}
%----------
where $\Lambda_{\pm}$ denotes the eigenvalue of the coefficient matrix of Eqs.~\eqref{eq:n_c2} and \eqref{eq:n_e2};
%----------
% Equation
%----------
\begin{align*}
    \Lambda_{\pm} =& -\frac{\gamma+\kappa+R_{+,-}+R_{-,+}}{2} \nonumber \\ 
    &\pm \sqrt{ \left( \frac{\kappa-\gamma+R_{+,-}-R_{-,+}}{2} \right)^2 + R_{+,+}R_{-,-} }.
\end{align*}
%----------
By assuming $\kappa \gg \gamma$, the first term in Eq.~\eqref{eq:ne(t)} dominates the evolution of $n_{\e}(t)$.
As a result, the transition rate, $W$, of the initially excited TLS can be obtained as 
%----------
% Equation
%----------
\begin{align}
    W \simeq & -\Lambda_{+} \nonumber\\
    =& \frac{\gamma+\kappa+R_{+,-}+R_{-,+}}{2} \nonumber \\ 
    &- \sqrt{ \left( \frac{\kappa-\gamma+R_{+,-}-R_{-,+}}{2} \right)^2 + R_{+,+}R_{-,-} }.
    \label{eq:W}
\end{align}
%----------
We note that $W$ is described by $-\Lambda_{-}$, instead, for $\kappa \ll \gamma$.
However, we restrict our discussion to the case for $\kappa \gg \gamma$, for simplicity, in accordance with Fano's approach~\cite{Fano61}.
Eq.~\eqref{eq:W} is one of our main results.

%----------
% Figure
%----------
\begin{figure*}
\centering
\includegraphics[width=.98\textwidth]{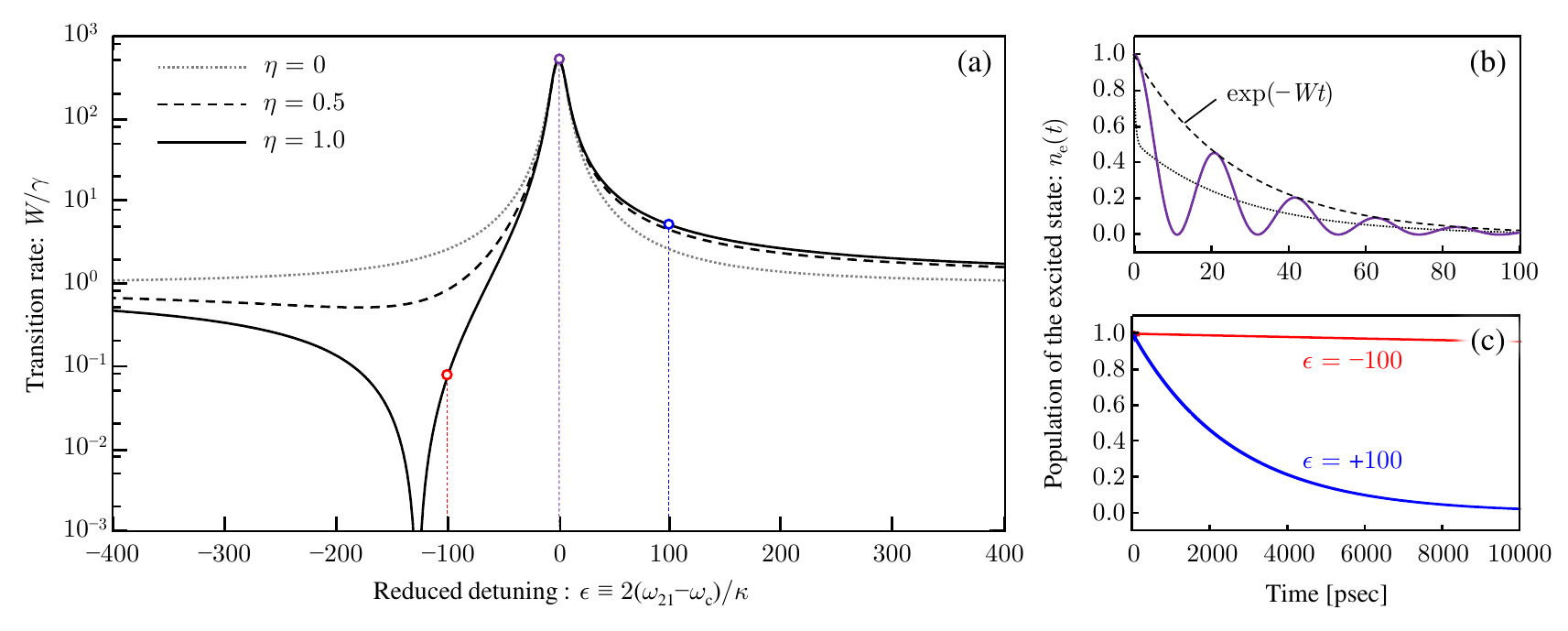}
\caption{
    Numerical results for $\gamma_{\ph} = 0$.
    (a) The transition rate, $W$, as a function of the reduced detuning.
    The dotted and solid lines, respectively, indicate the orthogonal ($\eta=0$) and identical ($\eta=1$) spatial radiation patterns of the TLS and the cavity.
    The dashed line is the result for the intermediate degree of overlap ($\eta=0.5$).
    (b) The time evolution of the population, $n_{\e}(t)$, under the on-resonance condition $\eps = 0$ with $\eta=1$.
    The solid line is obtained from Eqs.~\eqref{eq:n_c}--\eqref{eq:p}.
    The oscillating behavior means that the system is in the strong coupling regime.
    For comparison, the dotted line shows Eq.~\eqref{eq:ne(t)}, while the dashed line gives $\exp(-Wt)$ with Eq.~\eqref{eq:W}.
    (c) The same as (b) but under the off-resonance conditions, $\eps = -100$ (the red line) and $\eps = +100$ (the blue line).
    In this panel, the results by Eq.~\eqref{eq:ne(t)} and $\exp(-Wt)$ are not shown for the sake of visibility because these are almost exactly overlapped with the presented lines.
    The open circles in panel (a) correspond to the time evolutions in panels (b) and (c).
    The parameters are assumed $\kappa=50~\mu$eV, $\gamma=0.05~\mu$eV, $|g|=100~\mu$eV, and $\theta_{\c} = 0$.
}
\label{fig:W}
\end{figure*}
%----------

In order to discuss the meaning of our analysis and the obtained result [Eq.~\eqref{eq:W}], we show typical numerical results with no pure dephasing ($\gamma_{\ph}=0$) in Fig.~\ref{fig:W}.
In our calculations, other parameters are assumed $\kappa=50~\mu\mathrm{eV}$, $\gamma=0.05~\mu\mathrm{eV}$, $|g|=100~\mu$eV, and $\theta_{\c} = 0$~\cite{Hennessy07,Winger09,Ota15}, unless otherwise stated.
In Fig.~\ref{fig:W}(a), we can see that the transition rate, $W$, shows symmetric profile as a function of the reduced detuning $\eps \equiv 2(\omega_{21}-\omega_{\c})/\kappa$ when the spatial radiation patterns of the TLS and the cavity are orthogonal, $\eta=0$ (the dotted line).
This profile means that the transition rate is simply enhanced when the TLS comes into resonance with the cavity, as expected.
In contrast, the profile becomes asymmetric with increasing the value of $\eta$, i.e., the degree of overlap between the spatial radiation patterns.
This is a characteristic signature of the Fano effect.

For the identical radiation patterns, $\eta = 1$, we also show the time evolutions obtained by Eqs.~\eqref{eq:n_c}--\eqref{eq:p} under the on-resonant and off-resonant conditions in Figs.~\ref{fig:W}(b) and \ref{fig:W}(c), respectively.
In Fig.~\ref{fig:W}(b), we can find an oscillating behavior in the population of the excited state, $n_{\e}(t)$.
This corresponds to the Rabi oscillation, indicating that the system is in the strong coupling regime.
This is consistent with our setting of the parameters, $|g| \gtrsim \kappa + \gamma$.
We note that this parameter regime is beyond the applicable range of the Fano formula [Eq.~\eqref{eq:Fano}], despite the asymmetric profile of the transition rate.

To check the validity of our analysis, therefore, we also show the time evolutions by Eqs.~\eqref{eq:ne(t)} and \eqref{eq:nc(t)} (the dotted line) and by $\exp(-Wt)$ with Eq.~\eqref{eq:W} (the dashed line) in Fig.~\ref{fig:W}(b).
Here, we can see that the dotted line is roughly along the center line of the oscillations (the coarse-grained evolution), and as a result, the dashed line shows good agreement with the decay of the envelope.
These results suggest that our approach works well because, in the strong coupling regime, the transition rate is measured by the decay of its envelope by using the coarse-grained evolution, as mentioned above.
In Fig.~\ref{fig:W}(c), we can also verify that the asymmetric transition rate in Fig.~\ref{fig:W}(a) indeed gives a difference in the time evolutions between the positive and negative detuning.
We note that the results by Eqs.~\eqref{eq:ne(t)} and \eqref{eq:nc(t)} and by $\exp(-Wt)$ are not shown for the sake of visibility because these are almost exactly overlapped with the presented lines obtained by Eqs.~\eqref{eq:n_c}--\eqref{eq:p}.

We have thus obtained the formula of the transition rate $W$ [Eq.~\eqref{eq:W}] and numerically shown that the asymmetric profile can be still found even in the strong coupling regime.
However, it is important to show that our result can exactly recover the Fano formula.
For this purpose, we now focus on the weak coupling regime with $\kappa \gg \gamma$ in accordance with the perturbation approach by Fano~\cite{Fano61}.
In this case, the magnitude of the coupling constant $|g|$ is much smaller than $\gamma+\kappa \simeq \kappa$.
We can then neglect the term, $R_{+,+}R_{-,-}$, in the square root of Eq.~\eqref{eq:W} and obtain
%----------
% Equation
%----------
\begin{align}
    W \simeq \gamma + R_{-,+}.
    \label{eq:WCL}
\end{align}
%----------
It is interesting to note that, for $\eta = 0$ ($g_{\pm} = g$), Eq.~\eqref{eq:WCL} recovers the well-known Purcell effect $W \simeq \gamma + 2|g|^2 \frac{\Gamma_{\tot}}{\omega_{\c,21}^{2} + \Gamma_{\tot}^{2}}$
~\cite{Purcell46,Yamaguchi12}.
In cotrast, by assuming $\eta=1$ and $\gamma_{\ph}=0$, Eq.~\eqref{eq:WCL} yields
%----------
% Equation
%----------
\begin{align}
    W = \gamma + \gamma \Real\left[ \frac{(q-\ii)(q^{*}-\ii)}{1-\ii \eps} \right] = \gamma\frac{|q+\eps|^2}{\eps^{2}+1},
    \label{eq:FanoRecovery}
\end{align}
%----------
where $\eps = 2(\omega_{21}-\omega_{\c})/\kappa$ is the reduced detuning and $q$ is defined by
%----------
% Equation
%----------
\begin{align}
    q \equiv \frac{2|g|}{\sqrt{\gamma\kappa}}e^{\ii (\phi + \theta_{\c} - \theta_{21})}.
    \label{eq:q}
\end{align}
%----------
We note that Eq.~\eqref{eq:FanoRecovery} has the same form as the Fano formula [Eq.~\eqref{eq:Fano}], except that the parameter $q$ [Eq.~\eqref{eq:q}] is defined as a complex number in general.
This is consistent with the results discussed in charge transport experiments~\cite{Kobayashi02}.
In Eq.~\eqref{eq:q}, $\phi + \theta_{\c} -\theta_{21}$ denotes the phase difference between the direct channel, $\theta_{21}$, and the indirect channel via the discrete state (the cavity mode), $\phi + \theta_{\c}$.
As a result, by assuming that the parameter $q$ is real, the Fano formula [Eq.~\eqref{eq:Fano}] can be reproduced.
From these discussions, we can conclude that the presented dissipator, $\vD_{\F}$ [Eq.~\eqref{eq:FanoDissipator}],  indeed, can describe the Fano effect, and that the transition rate, $W$ [Eq.~\eqref{eq:W}], is a generalization of the Fano formula [Eq.~\eqref{eq:Fano}].

\begin{figure}
\centering
\includegraphics[width=.50\textwidth]{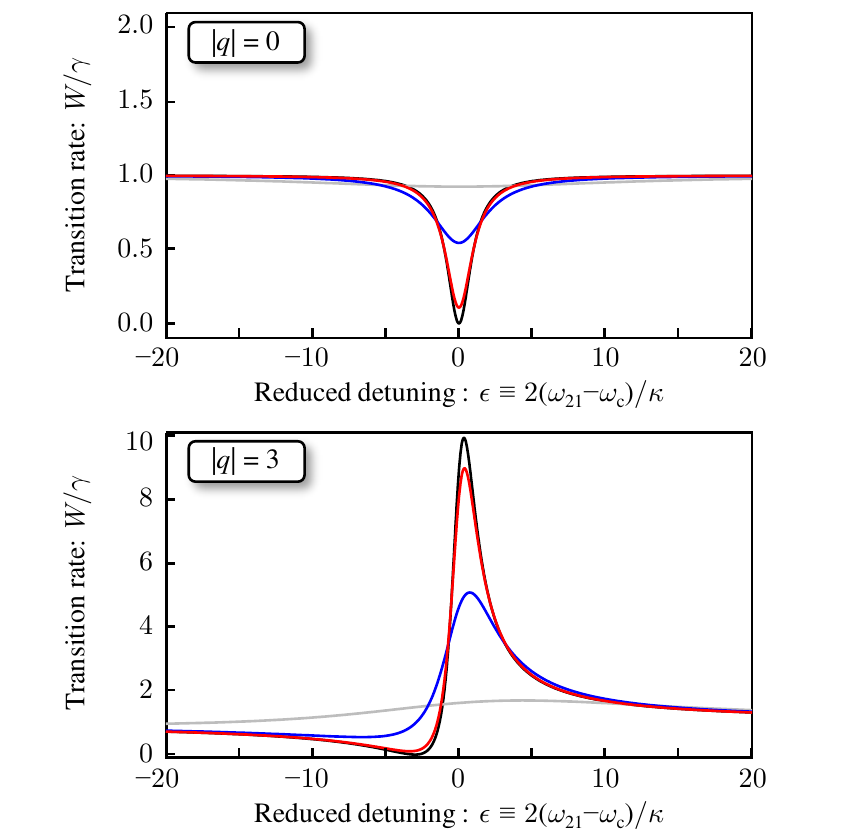}
\caption{
    The effect of pure dephasing on the transition rate; (a) $|q|=0$ ($|g|=0~\mu$eV) (b) $|q|=3$ ($|g|=2.37~\mu$eV).
    The pure dephasing rates, $\gamma_{\ph}$, are $0~\mu$eV (black line), $3~\mu$eV (red line), $30~\mu$eV (blue line), and $300~\mu$eV (gray line).
    The spatial radiation patterns are assumed identical between the TLS and the cavity ($\eta=1$).
    Other parameters are the same as Fig.~\ref{fig:W}.
}
\label{fig:Dephasing}
\end{figure}

Finally, we show the effect of pure dephasing for $|q|=0$ and $|q|=3$ in Figs.~\ref{fig:Dephasing}(a) and \ref{fig:Dephasing}(b), respectively.
In Fig.~\ref{fig:Dephasing}(a), we can find that the transition rate has a strong dip when the TLS is on resonance with the cavity for $\gamma_{\ph} = 0~\mu$eV (the black solid line).
This is consistent with the Fano formula.
In the context of cavity QED, however, this phenomena is interesting because $|q|=0$ means that there is no direct coupling between the TLS and the cavity mode, i.e., $g=0$ due to Eq.~\eqref{eq:q}.
As a result, at least in principle, an `antiresonance'~\cite{Miroshnichenko10} of the transition rate is possible by the destructive interference when $g=0$, while its resonance effect is often discussed as the Purcell effect.
By increasing the pure dephasing rate, however, we can see that the antiresonance effect is smeared out. 
Nevertheless, the effect of pure dephasing is weak up to $\simeq 3~\mu$eV.
This feature is generally understood from Eqs.~\eqref{eq:n_c}--\eqref{eq:p}; the pure dephasing does not strongly influence on the time evolution when $\gamma_{\ph} \ll (\gamma + \kappa)/2$ because $\gamma_{\ph}$ appears only in $\Gamma_{\tot}$ [Eq.~\eqref{eq:Gamma_tot}].
Hence, a similar behavior can be seen for $|g|=2.37~\ueV$ in Fig.~\ref{fig:Dephasing}(b), where the suppression of $W$ is again weakened by the pure dephasing, although the asymmetric profile can be still seen even if $\gamma_{\ph}$ becomes comparable to $(\gamma + \kappa)/2$.
The presented formulation thus allows us to quantitatively discuss the transition rate.
However, one may consider that the smeared profiles of $W$ are qualitatively trivial because, in general, the impact of interference is severely reduced by the pure dephasing; it is natural that the suppression of $W$ by the destructive interference goes away by the pure dephasing, for example.
Nevertheless, as we shall see in Sec.~\ref{subsec:Numerical results}, our study on the spectra shows that the interference responsible for the Fano effect is itself tolerant to the pure dephasing and that there is another reason for the incomplete suppression of the transition rate.

%----------
% Figure
%----------
%----------

\section{The Fano effect on the spectra}\label{sec:Spectra}
In the previous section, we have described a dissipator in the presence of the two direct and indirect dissipation channels, which allows us to study the physical quantities of the system with the Fano effect.
However, in reality, these quantities are usually measured via environmental degrees of freedom, i.e., the radiation modes in the continuum.
In this context, further analysis is required to discuss the observable quantities in addition to the simple modification of the QME.
In this section, we explain a formulation to obtain the emission spectra in a consistent manner with our treatment of the QME.
For this purpose, we first discuss the intensity detection in the presence of the Fano effect.
We then extend the idea to the spectroscopy.
Finally, we numerically show the Fano effect on the emission spectra.

\subsection{The intensity detection}\label{subsec:Intensity}
We here formulate the intensity, $S(t)$, evaluated as the total number of photons emitted from the system into all directions per unit time, as a preliminary step toward the emission spectra.
In general, for the intensity detection, the optical field from the system is introduced into a photon-counting detector after propagating a certain distance in the continuum.
Here, the incident energy per unit area and unit time on the detector, $\uI(\bm{r},t)$, is given by 
%----------
% Equation
%----------
\begin{align}
    \uI(\bm{r},t) = 2n\veps_{0}\c_{0}\uE^{-}(\bm{r},t)\cdot \uE^{+}(\bm{r},t),
    \label{eq:Intensity}
\end{align}
%----------
where $\bm{r}$ is the position of the detector and $n$, $\veps_{0}$ and $c_{0}$, respectively, denote the refractive index in the continuum, the vacuum dielectric constant, and the speed of light in vacuum.
$\breve{O}(t) \equiv e^{\ii\oH t}\oO e^{-\ii\oH t}$ indicates the Heisenberg picture for an arbitrary operator $\oO$ and the electromagnetic field operator, $\uE^{+}(\bm{r},t)$, in the continuum is described by
%----------
% Equation
%----------
\begin{align}
    \uE^{+}(\bm{r},t) = \sum_{\bm{k},\lambda}\bm{e}_{\bm{k},\lambda}\mathcal{E}_{\bm{k}}\ub_{\bm{k},\lambda}(t)e^{\ii \bm{k}\cdot\bm{r}},
    \label{eq:E+}
\end{align}
%----------
with 
%----------
% Equation
%----------
\begin{align}
    \mathcal{E}_{\bm{k}} \equiv \ii\sqrt{\frac{\omega_{\bm{k}}}{2n^2\veps_{0}V}}.
\end{align}
%----------
Here, $\bm{e}_{\bm{k},\lambda}$ is a unit vector along the polarization of the radiation mode ($\bm{k}$, $\lambda$) and $V$ is the quantization volume for the continuum.
We note that $\uE^{-}(\bm{r},t) = (\uE^{+}(\bm{r},t))^{\dagger}$.
Since $\uI(t) \equiv \int^{\pi}_{0}\dd \theta \int^{2\pi}_{0}\dd \phi r^{2}\sin\theta \uI(\bm{r},t)$ gives the power radiated from the system into all directions, $S(t)$ is given by $S(t) \simeq \Ex{\uI(t)}_{0}/\omega_{21} \simeq \Ex{\uI(t)}_{0}/\omega_{\c}$.
From the Heisenberg equation of motion, Eqs.~\eqref{eq:Htot}--\eqref{eq:HSB2} yield 
%----------
% Equation
%----------
\begin{align}
    \frac{\dd}{\dd t}\ub_{\bm{k},\lambda} = -\ii\omega_{\bm{k}}\ub_{\bm{k},\lambda} -\ii\xi^{*}_{\bm{k},\lambda}\usgm_{-} - \ii \zeta^{*}_{\bm{k},\lambda}\uac,
\end{align}
%----------
and its formal solution is given by
%----------
% Equation
%----------
\begin{align}
    \ub_{\bm{k},\lambda}(t) =& \ob_{\bm{k},\lambda}e^{-\ii\omega_{\bm{k}}t} -\ii\xi^{*}_{\bm{k},\lambda}\int^{t}_{0}\dd t' \usgm_{-}(t')e^{-\ii\omega_{\bm{k}}(t-t')} \nonumber \\
    &-\ii\zeta^{*}_{\bm{k},\lambda}\int^{t}_{0}\dd t'\uac(t')e^{-\ii\omega_{\bm{k}}(t-t')}.
    \label{eq:formal}
\end{align}
%----------
This equation means that the operator of the radiation modes, $\ub_{\bm{k},\lambda}$, is related to the two system operators, $\usgm_{-}$ and $\uac$.
By substituting Eq.~\eqref{eq:formal} into Eq.~\eqref{eq:E+}, $\uE^{+}(\bm{r},t)$ can be written as
%----------
% Equation
%----------
\begin{align}
    \uE^{+}(\bm{r},t) &= \uE^{+}_{\free}(\bm{r},t) + \uE^{+}_{\S}(\bm{r},t),
    \label{eq:Ef+ES}
\end{align}
%----------
where
%----------
% Equation
%----------
\begin{align}
    \uE^{+}_{\free}(\bm{r},t) \equiv \sum_{\bm{k},\lambda}\bm{e}_{\bm{k},\lambda}\mathcal{E}_{\bm{k},\lambda}\ob_{\bm{k},\lambda}e^{\ii (\bm{k}\cdot\bm{r}-\omega_{\bm{k}}t)}
\end{align}
%----------
denotes the free evolution of the radiation modes in the continuum and  
%----------
% Equation
%----------
\begin{align}
    \uE^{+}_{\S}(\bm{r},t) \equiv \uE^{+}_{21}(\bm{r},t) + \uE^{+}_{\c}(\bm{r},t),
    \label{eq:E21+Ec}
\end{align}
%----------
describes the electromagnetic field radiated from the system;
%----------
% Equation
%----------
\begin{align}
    \uE^{+}_{21}(\bm{r},t) &\equiv e^{-\ii\omega_{21}t} \int^{t}_{0}\dd t' \bm{G}_{21}(t,t')\tilde{\sigma}_{-}(t'), \\
    \uE^{+}_{\c}(\bm{r},t) &\equiv e^{-\ii\omega_{\c}t} \int^{t}_{0}\dd t' \bm{G}_{\c}(t,t')\tilde{a}_{\c}(t').
\end{align}
%----------
Here, $\tilde{\sigma}(t') \equiv \usgm(t')e^{\ii\omega_{21}t'}$ and $\tilde{a}_{\c}(t') \equiv \uac(t')e^{\ii\omega_{\c}t'}$ are the slowly varying operators~\cite{Scully97} and 
%----------
% Equation
%----------
\begin{align}
    \bm{G}_{21}(t,t') &\equiv -\ii\sum_{\bm{k},\lambda}\bm{e}_{\bm{k},\lambda}\mathcal{E}_{\bm{k}}\xi^{*}_{\bm{k},\lambda}
    e^{\ii\bm{k}\cdot\bm{r}-\ii\omega_{\bm{k},21}(t-t') }, 
    \label{eq:G21}\\
    \bm{G}_{\c}(t,t') &\equiv -\ii\sum_{\bm{k},\lambda}\bm{e}_{\bm{k},\lambda}\mathcal{E}_{\bm{k}}\zeta^{*}_{\bm{k},\lambda}
    e^{\ii\bm{k}\cdot\bm{r}-\ii\omega_{\bm{k},\c}(t-t') }
    \label{eq:Gc}
\end{align}
%----------
can be considered as a kind of propagator.
Therefore, by substituting Eqs.~\eqref{eq:Ef+ES} and \eqref{eq:E21+Ec} into Eq.~\eqref{eq:Intensity}, the expectation value of $\uI(\bm{r},t)$ is given by 
%----------
% Equation
%----------
\begin{align}
    I(\bm{r},t) =& 2n\eps_{0}c_{0} \Ex{\oE^{-}_{\S}(\bm{r}) \cdot \oE^{+}_{\S}(\bm{r})}_{t} \nonumber \\
                =& 2n\eps_{0}c_{0} \Ex{\oE^{-}_{21}(\bm{r}) \cdot \oE^{+}_{21}(\bm{r})}_{t} \nonumber \\
                &+ 2n\eps_{0}c_{0} \Ex{\oE^{-}_{\c}(\bm{r}) \cdot \oE^{+}_{\c}(\bm{r})}_{t} \nonumber \\
                &+ 4n\eps_{0}c_{0} \Real[\Ex{\oE^{-}_{21}(\bm{r}) \cdot \oE^{+}_{\c}(\bm{r})}_{t}],
    \label{eq:I(r,t)}
\end{align}
%----------
where $\Ex{\cdots \oE^{\pm}_{\free}(\bm{r}) \cdots}_{t} = 0$ has been used because the continuum is assumed in vacuum in our study.
In the right hand side of the second equality, the first and second terms result from the radiation from the TLS and the cavity, respectively, whereas the last term indicates their interference. 

In order to study $S(t)$, therefore, further analysis on $\uE^{+}_{21}(\bm{r},t)$ and $\uE^{+}_{\c}(\bm{r},t)$ is required.
For this purpose, we introduce the following approximations for Eqs.~\eqref{eq:G21} and \eqref{eq:Gc},
%----------
% Equation
%----------
\begin{align}
    \bm{G}_{21}(t,t') &\simeq -\ii\sqrt{2}\bar{\xi}^{*}\bm{e}_{21}(\bm{r})\sum_{\bm{k}}\mathcal{E}_{\bm{k}}
    e^{\ii\bm{k}\cdot\bm{r}-\ii\omega_{\bm{k},21}(t-t') }, \\
    \bm{G}_{\c}(t,t') &\simeq -\ii \sqrt{2}\bar{\zeta}^{*}\bm{e}_{\c}(\bm{r}) \sum_{\bm{k}}\mathcal{E}_{\bm{k}}e^{\ii\bm{k}\cdot\bm{r}-\ii\omega_{\bm{k},\c}(t-t') },
\end{align}
%----------
based on the assumptions in Sec.~\ref{sec:Setup}.
Here, $\bm{e}_{21}(\bm{r})$ and $\bm{e}_{\c}(\bm{r})$ are the unit vectors along the polarization of the radiation patterns from the TLS and the cavity, respectively, in the far field region ($kr \gg 1$).
A value of $\sqrt{2}$ results from the sum of the two orthogonal unit vectors along the polarization.
Then, we can calculate
%----------
% Equation
%----------
\begin{align}
    &\sum_{\bm{k}} \mathcal{E}_{\bm{k}} e^{\ii\bm{k}\cdot\bm{r} - \ii\omega_{\bm{k},\alpha}(t-t')} \nonumber \\
        &= \frac{V}{(2\pi)^{3}}\int^{\infty}_{0}\dd k \int^{2\pi}_{0} \dd\phi_{k} \int^{\pi}_{0} \dd\theta_{k} k^2 \sin\theta_{k} \nonumber \\
        &\quad \times \mathcal{E}_{k} e^{\ii kr \cos\theta_{k} - \ii \omega_{k,\alpha}(t-t')} \nonumber \\
        &\simeq -\ii\frac{\pi c_{0} \mathcal{E}_{\alpha}}{2nr\omega_{\alpha}} D(\omega_{\alpha}) \nonumber \\
        &\quad \times \{ \delta(t-t'-t_{\d})e^{\ii\omega_{\alpha}t_{\d}} - \delta(t-t'+ t_{\d})e^{-\ii\omega_{\alpha}t_{\d}} \},
\end{align}
%----------
where $t_{\d} \equiv nr/c_{0}$ is the delay time due to the propagation of light from the origin, $D(\omega) = \frac{n^{3}\omega^{2}V}{\pi^{2} c^{3}_{0}}$, is the density of states in the free space, and $\alpha \in \{21, \c \}$.
Therefore, we have
%----------
% Equation
%----------
\begin{align}
    \uE^{+}_{21}(\bm{r},t) &\simeq -\bm{e}_{21}(\bm{r}) \frac{\pi\mathcal{E}_{21}c_{0}}{\sqrt{2}nr\omega_{21}}\bar{\xi}^{*} D(\omega_{21})\usgm_{-}(t-t_{\d}),
    \label{eq:E21} \\
    \uE^{+}_{\c}(\bm{r},t) &\simeq -\bm{e}_{\c}(\bm{r}) \frac{\pi\mathcal{E}_{\c}c_{0}}{\sqrt{2}nr\omega_{\c}}\bar{\zeta}^{*} D(\omega_{\c})\uac(t-t_{\d}).
    \label{eq:Ec}
\end{align}
%----------
By substituting these equations into Eq.~\eqref{eq:I(r,t)} with neglecting the delay time, $t_{\d}$, for simplicity, we can obtain
%----------
% Equation
%----------
\begin{align}
    S(t) \simeq & \gamma \Ex{\osgm_{+}\osgm_{-}}_{t} + \kappa \Ex{\oacd\oac}_{t} +2\Real[\gamma_{\F}\Ex{\osgm_{+}\oac}_{t}],
    \label{eq:S(t)}
\end{align}
%----------
where, based on the spirit of the phenomenological parameter $\eta$ in Eq.~\eqref{eq:gamma_F}, we have estimated 
%----------
% Equation
%----------
\begin{align*}
    e^{\ii(\theta_{21}-\theta_{\c})}\sqrt{\gamma\kappa} \int^{\pi}_{0}\dd\theta \int^{2\pi}_{0}\dd\phi \frac{\bm{e}_{21}(\bm{r})\cdot\bm{e}_{\c}(\bm{r})\sin\theta}{4\pi} \simeq \gamma_{\F},
\end{align*}
%----------
because the integration means the average over the entire solid angle.
It is interesting to note that, in Eq.~\eqref{eq:S(t)}, the individual coefficients agree well with the dissipators in Sec.~\ref{subsec:QME}; the last term is the direct consequence of the Fano effect.
Thus, the effect of interference should be considered in the formulation of observable quantities as well as in the dissipators of the QME.

\subsection{The emission spectra}\label{subsec:Spectra}
For spectral detection, in general, the emitted photons are introduced into a spectral apparatus, in which the photons are dispersed by a spectrometer before the number of the photons is counted by a detector.
In this context, the spectrometer can be considered as a kind of a spectral filter for the detector.
Here, we assume that the characteristic function of this filtering is Lorentzian with a central frequency $\nu$ and a width $\ds$,
%----------
% Equation
%----------
\begin{align}
F(\omega) \equiv \frac{A}{(\ds/2)-\ii(\omega - \nu)},
\end{align}
%----------
where $A$ is a constant determined by the performance of the spectrometer.
Since the number of the photons incident on the detector depends on the central frequency $\nu$ and the width $\ds$, the average photon flux density on the detector at position $\bm{r}$ can be described as $S(\bm{r},t,\nu,\ds)$.
The emission spectra are then given by the profile of $S(\bm{r},t,\nu,\ds)$ as a function of $\nu$, in which $\ds$ corresponds to the spectral resolution~\cite{Eberly77}.
In the following, we discuss the spectra integrated over time and directions, $S(\nu,\ds) \equiv \int^{\infty}_{-\infty}\dd t \int^{\pi}_{0}\dd \theta \int^{2\pi}_{0}\dd \phi r^{2}\sin\theta S(\bm{r},t,\nu,\ds)$, for simplicity.

For this purpose, we describe the Fourier transform of the electromagnetic field by
%----------
% Equation
%----------
\begin{align}
    \uE^{+}(\bm{r},\omega) \equiv \int^{\infty}_{-\infty} \dd t' \uE^{+}(\bm{r},t') e^{\ii\omega t'}.
\end{align}
%----------
Since $\uE^{+}(\bm{r},\omega)$ is composed of the free evolution of the radiation modes and the filtered field from the system,
%----------
% Equation
%----------
\begin{align}
    \uE^{+}(\bm{r},\omega) = \uE^{+}_{\free}(\bm{r},\omega) + F(\omega)\uE^{+}_{\S}(\bm{r},\omega),
\end{align}
%----------
the inverse Fourier transform yields
%----------
% Equation
%----------
\begin{align}
    \uE^{+}(\bm{r},t) = \uE^{+}_{\free}(\bm{r},t) + \int^{\infty}_{-\infty}\dd t'  F(t-t')\uE^{+}_{\S}(\bm{r},t'),
\end{align}
%----------
where $F(t) = Ae^{-(\ds/2+\ii\nu)t}\Theta(t)$ and $\Theta(t)$ is the step function.
Hence, in a similar manner to Eqs.~\eqref{eq:Intensity} and \eqref{eq:I(r,t)},
the expectation value of the incident energy per unit area and unit time on the detector is given by
%----------
% Equation
%----------
\begin{align}
    &I(\bm{r},t,\nu,\ds) = 2n\eps_{0}c_{0} \int^{\infty}_{-\infty}\dd t_{1} \int^{\infty}_{-\infty}\dd t_{2}\nonumber\\
    &\qquad \times F^{*}(t-t_{1})F(t-t_{2})\Ex{\uE^{-}_{\S}(\bm{r},t_{1})\cdot\uE^{+}_{\S}(\bm{r},t_{2})}_{0} \nonumber \\
    &\quad = 2\pi|A|^{2}\int^{\infty}_{0}\dd t' J(\bm{r},t-t',\nu,\ds)e^{-\ds t'},
    \label{eq:I(r,t,nu,ds)}
\end{align}
%----------
where we have defined 
%----------
% Equation
%----------
\begin{align*}
    &J(\bm{r},t,\nu,\ds) \equiv \frac{2n\eps_{0}c_{0}}{\pi} \nonumber \\
    & \quad \times \Real{ \int^{\infty}_{0}\dd\tau 
    \Ex{\uE^{-}_{\S}(\bm{r},t-\tau)\cdot\uE^{+}_{\S}(\bm{r},t)}_{0} e^{(\ii\nu-\ds/2)\tau} }.
\end{align*}
%----------
In the second equality of Eq.~\eqref{eq:I(r,t,nu,ds)}, we have used
%----------
% Equation
%----------
\begin{align*}
    &{\textstyle \int^{t}_{-\infty}\dd t_{1}\int^{t}_{-\infty}\dd t_{2}[\cdots]}  \\
      &\qquad = {\textstyle \int^{\infty}_{0}\dd t'_{1}\int^{\infty}_{0}\dd t'_{2} [\cdots] 
      + \int^{0}_{-\infty}\dd t'_{1} \int^{\infty}_{-\tau}\dd t'_{2} [\cdots]},
\end{align*}
%----------
by a transformation of variables, $t'_{1} = t_{1} - t_{2}$ and $t'_{2}=t-t_{1}$.
Since $I(\bm{r},t,\nu,\ds)$ depends on the coefficient $A$, we here set a condition of normalization,
%----------
% Equation
%----------
\begin{align*}
    \int^{\infty}_{-\infty}\dd t \int^{\infty}_{-\infty} \dd \nu \frac{I(\bm{r},t,\nu,\ds)}{2n\eps_{0}c_{0}}
     = \int^{\infty}_{-\infty}\dd t \Ex{\oE^{-}_{\S}(\bm{r}) \cdot \oE^{+}_{\S}(\bm{r})}_{t}.
\end{align*}
%----------
This equation means that the integration of $I(\bm{r},t,\nu,\ds)$ over frequency $\nu$ and time $t$ is identical to the total energy detected at position $\bm{r}$.
We can then obtain $2\pi|A|^{2} = \ds$ from Parseval's theorem~\cite{Eberly77}.
Hence, by integrating Eq.~\eqref{eq:I(r,t,nu,ds)} over all directions and time, $S(\nu,\ds) \simeq I(\nu,\ds)/\omega_{21} \simeq I(\nu,\ds)/\omega_{\c}$ yields
%----------
% Equation
%----------
\begin{align}
    S(\nu,\ds) = S_{21}(\nu,\ds) + S_{\c}(\nu,\ds) + S_{\F}(\nu,\ds),
    \label{eq:S(nu)}
\end{align}
%----------
with
%----------
% Equation
%----------
\begin{align*}
    S_{21}(\nu,\ds) &= \frac{\gamma}{\pi} \Real {\textstyle \int^{\infty}_{-\infty}\dd t \int^{\infty}_{0}\dd \tau} \Ex{\usgm_{+}(t-\tau)\usgm_{-}(t)}_{0} e^{(\ii\nu-\frac{\ds}{2})\tau}, \nonumber \\
    S_{\c}(\nu,\ds) &= \frac{\kappa}{\pi} \Real {\textstyle \int^{\infty}_{-\infty}\dd t \int^{\infty}_{0}\dd \tau} \Ex{\uacd(t-\tau)\uac(t)}_{0} e^{(\ii\nu-\frac{\ds}{2})\tau}, \nonumber \\
    S_{\F}(\nu,\ds) &= \frac{1}{\pi} \Real {\textstyle \int^{\infty}_{-\infty}\dd t \int^{\infty}_{0}\dd \tau} 
    \left( \gamma_{\F}\Ex{\usgm_{+}(t-\tau)\uac(t)}_{0} \right. \nonumber\\
    &\quad \left. + \gamma^{*}_{\F}\Ex{\uacd(t-\tau)\usgm_{-}(t)}_{0} \right) e^{(\ii\nu-\frac{\ds}{2})\tau},
\end{align*}
%----------
where we have used Eqs.~\eqref{eq:E21+Ec}, \eqref{eq:E21}, and \eqref{eq:Ec}.
$S_{21}(\nu,\ds)$ and $S_{\c}(\nu,\ds)$ are the time-integrated spectra of the TLS and the cavity, respectively.
In contrast, $S_{\F}(\nu,\ds)$ is the contribution due to their interference.
Again, the proportionality factors in these expressions agree well with the dissipators in Sec.~\ref{subsec:QME}.

For the application of this result, we note that the correlation functions of the form, $\Ex{\uO_{i}(t-\tau)\uO_{j}(t)}_{0}$, have to be evaluated.
However, the standard quantum regression theorem~\cite{Scully97,Carmichael99} cannot be directly applied because its form is simply outside the range of application.
To circumvent this difficulty, we first consider $\Ex{\uO_{i}(t)\uO_{j}(t+\tau)}_{0}$.
The quantum regression theorem, then, allows us to express the correlation function in the form; 
$\Ex{\uO_{i}(t)\uO_{j}(t+\tau)}_{0}= \sum_{k}C_{jk}(\tau)\Ex{\oO_{i}\oO_{k}}_{t}$ when $\tau \ge 0$.
By substituting $t \rightarrow t-\tau$, therefore, we can obtain $\Ex{\uO_{i}(t-\tau)\uO_{j}(t)}_{0}= \sum_{k}C_{jk}(\tau)\Ex{\oO_{i}\oO_{k}}_{t-\tau}$.
As a result, we can estimate $\Ex{\uO_{i}(t-\tau)\uO_{j}(t)}_{0}$ by appling the quantum regression theorem, in which the dissipator, $\vD_{\F}$, again plays a key role to evaluate the correlation functions.
In this sense, we note that the consistency between the treatments of the QME and the spectra has great importance.

\begin{widetext}
By assuming the initially excited TLS, then, we obtain 
%----------
% Equation
%----------
\begin{align}
    S_{\alpha}(\nu,\ds) = \Real\left[\frac{1}{\gamma_{+}-\gamma_{-}}\left(\frac{f_{\alpha}(\gamma_{+})}{\ii\nu+\gamma_{+}-\ds/2} - \frac{f_{\alpha}(\gamma_{-})}{\ii\nu+\gamma_{-}-\ds/2} \right)\right]
    \qquad {\text{for }} \alpha \in \{21,\c,\F \},
    \label{eq:Sa(nu)}
\end{align}
%----------
where we have defined
%----------
% Equation
%----------
\begin{align*}
    &f_{21}(\gamma_{\pm}) \equiv \frac{\gamma}{\pi}\left\{ \ii g_{-}\mI_{p} - \left( \gamma_{\pm} + \ii\omega_{\c} + \frac{\kappa}{2} \right) \mI_{\e} \right\}, \qquad
    f_{\c}(\gamma_{\pm}) \equiv \frac{\kappa}{\pi}\left\{ \ii g^{*}_{+}\mI^{*}_{p} - \left( \gamma_{\pm} + \ii\omega_{21} + \gamma_{\ph} + \frac{\gamma}{2} \right) \mI_{\c} \right\}, \\
    &f_{\F}(\gamma_{\pm}) \equiv \frac{1}{\pi}\left\{ \ii g_{-} \gamma^{*}_{\F}\mI_{\c} + \ii g^{*}_{+}\gamma_{\F}\mI_{\e}
     -\left( \gamma_{\pm} + \ii\omega_{\c} + \frac{\kappa}{2} \right)\gamma^{*}_{\F}\mI^{*}_{p}
     -\left( \gamma_{\pm} + \ii\omega_{21} + \gamma_{\ph} + \frac{\gamma}{2} \right) \gamma_{\F}\mI_{p} \right\},
\end{align*}
%----------
with $\mI_{\e} = \int^{\infty}_{-\infty} n_{\e}(t) \dd t$, $\mI_{\c} = \int^{\infty}_{-\infty} n_{\c}(t) \dd t$, and $\mI_{p} = \int^{\infty}_{-\infty} p(t) \dd t$.
In addition, $\gamma_{\pm}$ is given by
%----------
% Equation
%----------
\begin{align}
    \gamma_{\pm} \equiv -\frac{1}{2}\left( \Gamma_{\tot} + \ii(\omega_{21}+\omega_{\c}) \right) 
    \pm \frac{1}{2}\sqrt{\left(\frac{\kappa-\gamma}{2}-\gamma_{\ph} + \ii\omega_{\c,21} \right)^2 - 4g^{*}_{+}g_{-}}.
\end{align}
%----------
\end{widetext}
These expressions are the second of our main results in our theoretical treatments.
Here, we note that the obtained spectrum, $S(\nu,\ds)$, has an explicit physical meaning, i.e., the number of photons counted per unit frequency, by definition.
Therefore, the integration of the total spectrum over the frequency, $\int^{\infty}_{-\infty}\dd\nu S(\nu,\ds)$, indicates the total number of photons finally emitted from the system, which must be one because we consider the initially excited TLS and its relaxation within the linear optical process.
In order to verify this prediction, we analytically integrate $S(\nu,\ds)$ and obtain
%----------
% Equation
%----------
\begin{align}
    \int^{\infty}_{-\infty}\dd\nu S(\nu,\ds) = \gamma\mI_{\e} + \kappa\mI_{\c} + 2 \Real [\gamma_{\F}\mI_{p}],
    \label{eq:nu-integration}
\end{align}
%----------
where
%----------
% Equation
%----------
\begin{align*}
    \int^{\infty}_{-\infty}\dd \nu \frac{f_{\alpha}(\gamma_{\pm})}{\ii\nu+\gamma_{\pm}-\ds/2} = -\pi f_{\alpha}(\gamma_{\pm}).
\end{align*}
%----------
has been used.
We note that Eq.~\eqref{eq:nu-integration} is consistent with the time integration of Eq.~\eqref{eq:S(t)}.
Furthermore, by assuming $\lim_{t \rightarrow \infty} n_{\c}(t) = \lim_{t \rightarrow \infty} n_{\e}(t) = 0$, the time integrations of Eqs.~\eqref{eq:n_c} and \eqref{eq:n_e} together yield
%----------
% Equation
%----------
\begin{align*}
    n_{\c}(0)+n_{\e}(0) = \int^{\infty}_{0} \dd t (\gamma n_{\e} + \kappa n_{\c} + 2\Real (\gamma_{\F} p)).
\end{align*}
%----------
As a result, for $n_{\c}(0)=0$ and $n_{\e}(0)=1$, we can find
%----------
% Equation
%----------
\begin{align}
    \int^{\infty}_{-\infty}\dd\nu S(\nu,\ds) = 1,
    \label{eq:one}
\end{align}
%----------
where $n_{\c}(t)=n_{\e}(t)=p(t)=0$ for $t<0$ has been used.
Eq.~\eqref{eq:one} is consistent with the prediction that only one photon is finally emitted from the system and this result ensures the validity of our treatments, presented above.
It is obvious that the simultaneous consideration of $S_{\F}(\nu,\ds)$ and $\vD_{\F}$ is essential to achieve this property.

For our analysis on the spectra [Eq.~\eqref{eq:Sa(nu)}], in the next section (Sec.~\ref{subsec:Numerical results}), 
$\mI_{\e}$, $\mI_{\c}$, and $\mI_{p}$ are analytically evaluated by the coarse-grained evolution [Eqs.~\eqref{eq:Adiabatic}, \eqref{eq:ne(t)}, and \eqref{eq:nc(t)}], for simplicity.
We note that the validity of this approach can be checked by calculating Eq.~\eqref{eq:nu-integration} to be one.

%----------
% Figure
%----------
\begin{figure*}
\centering
\includegraphics[width=.98\textwidth]{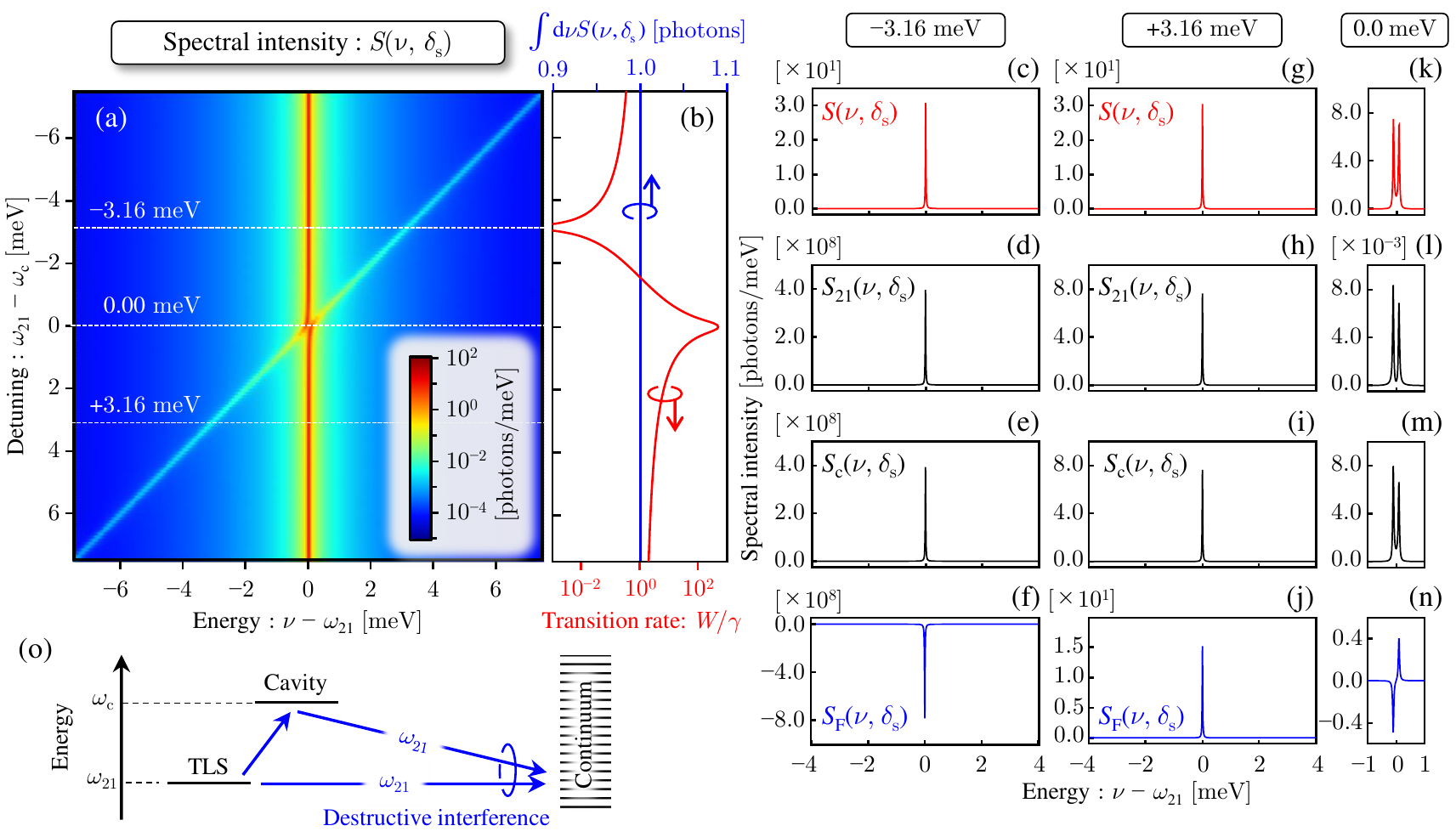}
\caption{
    Spectra for $\gamma_{\ph} = 0$ when the spatial radiation patterns of the TLS and the cavity are identical ($\eta=1$).
    The resolution of the spectrometer is set to be $\ds = 20~\ueV$.
    Other parameters are the same as in Fig.~\ref{fig:W}.
    (a) The dependence of the spectral intensity, $S(\nu,\ds)$, on the detuning.
    (b) The transition rate, $W/\gamma$, and the integrated value, $\int \dd \nu S(\nu,\ds)$, as a function of the detuning.
    We note that $W/\gamma$ is identical to the solid line in Fig.~\ref{fig:W}(a).
    $\int \dd\nu S(\nu,\ds)$ is calculated by Eq.~\eqref{eq:nu-integration} to check the validity of our numerical results.
    (c)--(n) The decomposition of $S(\nu,\ds)$ [panels (c), (g), (k)] into $S_{21}(\nu,\ds)$ [panels (d), (h), (l)], $S_{\c}(\nu,\ds)$ [panels (e), (i), (m)], and $S_{\F}(\nu,\ds)$ [panels (f), (j), (n)].
    The detunings are $-3.16~\meV$ for the left column [panels (c)--(f)], $+3.16~\meV$ for the middle column [panels (g)--(j)], and $0.0~\meV$ for the right column [panels (k)--(n)].
    These detunings are indicated by the dashed lines in panel~(a).
    (o) A schematic illustration of the dissipation channels for $\omega_{21}<\omega_{\c}$.
    We note that the frequency of the field escaped via the cavity is dominated by $\omega_{21}$, instead of $\omega_{\c}$.
    This process is achieved via the virtual photon excitation inside the cavity and is the same as the classical forced oscillation.
}
\label{fig:SpWithoutDephasing}
\end{figure*}
%----------

\subsection{Numerical results}\label{subsec:Numerical results}
Figure~\ref{fig:SpWithoutDephasing} shows numerical results for $\gamma_{\ph}=0$ with $\eta=1$, where the resolution of the spectrometer is set to be $\ds = 20~\mu$eV.
Other parameters are the same as in Fig.~\ref{fig:W}.
In Fig.~\ref{fig:SpWithoutDephasing}(a), we can see that the spectra, $S(\nu,\ds)$, are almost symmetric with respect to $\nu - \omega_{21}$ with changing the sign of the detuning $\omega_{21}-\omega_{\c}$.
This is in contrast to the asymmetric profile of the transition rate, $W$, shown in Fig.~\ref{fig:SpWithoutDephasing}(b).
The validity of these numerical results is supported by $\int^{\infty}_{-\infty}\dd\nu S(\nu,\ds) \simeq 1$ over the entire range of the calculations, as shown in Fig.~\ref{fig:SpWithoutDephasing}(b).
Furthermore, the color map is nearly the same as the spectra for $\eta = 0.0$ although we do not show the results here.
Nevertheless, as we explain below, the Fano effect plays an essential role for a consistent understanding of these results.

To elucidate these points, at positive and negative detunings, $\omega_{21}-\omega_{\c}=\pm 3.16~\meV$, the spectra are decomposed into $S_{21}(\nu,\ds)$, $S_{\c}(\nu,\ds)$, and $S_{\F}(\nu,\ds)$, as shown in Figs.~\ref{fig:SpWithoutDephasing}(c)--(j).
In the total spectra, $S(\nu,\ds)$, for both detunings [Figs.~\ref{fig:SpWithoutDephasing}(c) and \ref{fig:SpWithoutDephasing}(g)], the main peaks appear at the TLS transition energy, $\omega_{21}$, with comparable spectral intensities.
However, $S_{21}(\nu,\ds)$ for $-3.16~\meV$ detuning [Fig.~\ref{fig:SpWithoutDephasing}(d)] is a factor of $10^7 \sim 10^8$ larger than that for $+3.16~\meV$ detuning [Fig.~\ref{fig:SpWithoutDephasing}(h)].
A similar behavior can also be seen in $S_{\c}(\nu,\ds)$ [Figs.~\ref{fig:SpWithoutDephasing}(e) and \ref{fig:SpWithoutDephasing}(i)].
This means that the spectra are strongly asymmetric if we just consider $\vD_{\F}$ in the QME but without the contribution of $S_{\F}(\nu,\ds)$.
However, the spectral intensities are largely reduced by a destructive interference for $-3.16~\meV$ detuning [Fig.~\ref{fig:SpWithoutDephasing}(f)], while enhanced by a constructive interference for $+3.16~\meV$ detuning [Fig.~\ref{fig:SpWithoutDephasing}(j)].
As a result, $S(\nu,\ds)$ gives a similar degree of spectral intensities for $\pm3.16~\meV$ detunings.
In contrast, at resonance condition ($\omega_{21}-\omega_{\c}=0~\meV$), an asymmetric vacuum Rabi splitting can be obtained, as a consequence of the interference [Figs.~\ref{fig:SpWithoutDephasing}(k)--(n)].
This result is in agreement with Ref.~\cite{Ota15}.

The Fano effect is thus essential for a consistent understanding of the emission spectra.
However, as stated above, there is a remarkable similarity between the total spectra for $\eta = 1.0$ and $\eta = 0.0$ (not shown) although the transition rate, $W$, depends significantly on the value of $\eta$ [Fig.~\ref{fig:W}(a)].
This is because the TLS has no other choice but to finally emit a photon at the transition energy, $\omega_{21}$, after an infinite time, even in the presence of the interference.
This situation is schematically illustrated in Fig.~\ref{fig:SpWithoutDephasing}(o) by assuming $\omega_{21}<\omega_{\c}$, where we note that the frequency of the field escaped via the cavity is dominated by $\omega_{21}$, instead of $\omega_{\c}$.
This feature is evidenced by the dominant contribution of the emission line at $\nu = \omega_{21}$ in $S_{\c}(\nu,\ds)$ [Fig.\ref{fig:SpWithoutDephasing}(e)].
Such a transition process is mediated via the virtual photon excitation inside the cavity and can be interpreted in the same manner as the classical forced oscillation.
Therefore, the direct and indirect dissipation channels both produce fields with frequency $\omega_{21}$, as seen in Fig.~\ref{fig:SpWithoutDephasing}(o).
In consequence, the TLS has to finally emit a photon at its transition energy, $\omega_{21}$, regardless of the interference.
Although the spectral width can be changed due to the modification of the transition rate, the difference is much below the resolution ($\ds = 20~\ueV$).
As a result, the total spectra cannot be significantly modified by the Fano effect alone by considering that the $\nu$-integral of $S(\nu,\ds)$ is fixed to one in the presented situation.

%----------
% Figure
%----------
\begin{figure*}
\centering
\includegraphics[width=.98\textwidth]{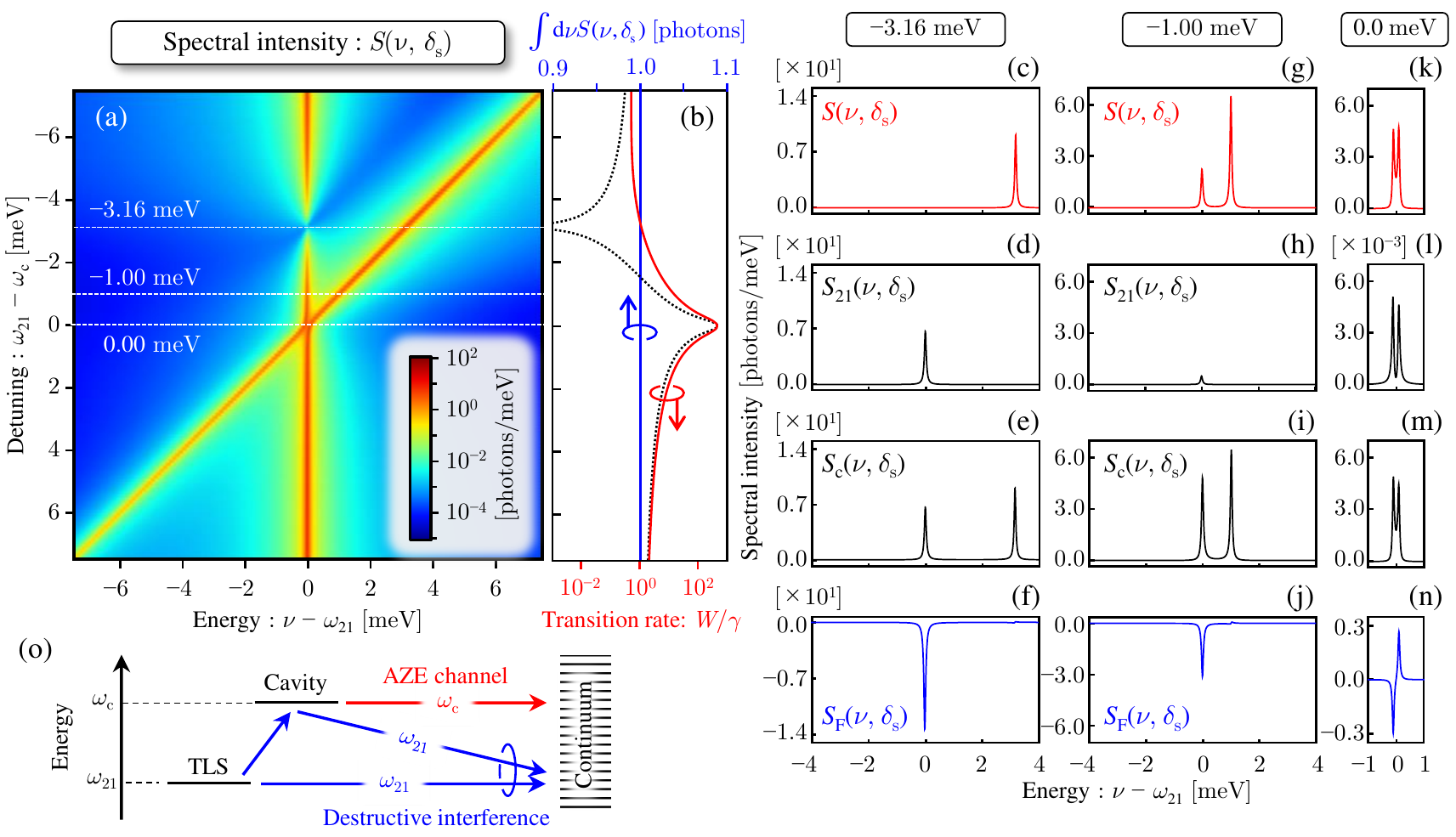}
\caption{
    Spectra for $\gamma_{\ph} = 30.0~\ueV$.
    Other parameters are the same as in Fig.~\ref{fig:SpWithoutDephasing}.
    (a) The dependence of the spectral intensity, $S(\nu,\ds)$, on the detuning.
    (b) The transition rate, $W/\gamma$, and the integrated value, $\int \dd\nu S(\nu,\ds)$ [Eq.~\eqref{eq:nu-integration}], as a function of the detuning.
    For comparison, the transition rate for $\gamma_{\ph} = 0.0~\ueV$ is indicated by the dotted line, which is identical to the solid line in Fig.~\ref{fig:W}(a).
    (c)--(n) The decomposition of $S(\nu,\ds)$ into $S_{21}(\nu,\ds)$, $S_{\c}(\nu,\ds)$, and $S_{\F}(\nu,\ds)$, in a similar manner to Figs.~\ref{fig:SpWithoutDephasing}(c)--(n).
    The detunings are $-3.16~\meV$ for the left column [panels (c)--(f)], $-1.00~\meV$ for the middle column [panels (g)--(j)], and $0.0~\meV$ for the right column [panels (k)--(n)].
    These detunings are again indicated by the dashed lines in panel~(a).
    (o) A schematic illustration of the dissipation channels for $\omega_{21}<\omega_{\c}$.
    The AZE channel is additionally opened by the pure dephasing.
}
\label{fig:SpWithDephasing}
\end{figure*}
%----------

However, the situation is drastically changed when the pure dephasing is additionally introduced, as shown in Fig.~\ref{fig:SpWithDephasing}, where the parameters are the same as in Fig.~\ref{fig:SpWithoutDephasing} except that $\gamma_{\ph}=30.0~\ueV$.
In Fig.~\ref{fig:SpWithDephasing}(a), the spectra, $S(\nu,\ds)$, show an asymmetric behavior in contrast to Fig.~\ref{fig:SpWithoutDephasing}(a).
For positive detuning ($\omega_{21}>\omega_{\c}$), the intensity at the cavity resonance is enhanced in comparison with Fig.~\ref{fig:SpWithoutDephasing}(a).
This effect can be understood from the viewpoint of the quantum anti-Zeno effect (AZE)~\cite{Yamaguchi12,Yamaguchi08-2}, known as one of the mechanisms responsible for the off-resonant cavity feeding~\cite{Yamaguchi12,Yamaguchi08-2,Ates09,Suffczynski09,Jarlov16}.
For negative detuning ($\omega_{21}<\omega_{\c}$), in contrast, we can find a peculiar spectral behavior.
At the detuning of $-3.16~\meV$, in particular, a strong intensity reduction at $\nu = \omega_{21}$ can be seen with a further intensity enhancement at the cavity resonance ($\nu=\omega_{\c}$).
This value of detuning corresponds to the original detuning position achieving the minimum of the transition rate for $\gamma_{\ph}=0$, instead of the shifted position for $\gamma_{\ph}=30~\ueV$ [Fig.~\ref{fig:SpWithDephasing}(b)].
Especially for negative detuning, thus, the Fano effect largely influences on the spectra with the help of pure dephasing, even if the transition rate is significantly washed out ($\gamma_{\ph} > (\gamma+\kappa)/2$).
This is seemingly counterintuitive if we consider that the pure dephasing spoils the interference between the two dissipation channels.
Again, $\int^{\infty}_{-\infty}\dd\nu S(\nu,\ds) \simeq 1$ in Fig.~\ref{fig:SpWithDephasing}(b) ensures the validity of these results.

To clarify the underlying physics, at the detuning of $-3.16~\meV$, the total spectrum is again decomposed into $S_{21}(\nu,\ds)$, $S_{\c}(\nu,\ds)$, and $S_{\F}(\nu,\ds)$, in Figs.~\ref{fig:SpWithDephasing}(c)--(f).
As seen in Fig.~\ref{fig:SpWithDephasing}(a), $S(\nu,\ds)$ has a single main peak at the cavity resonance, $\omega_{\c}$, in Fig.~\ref{fig:SpWithDephasing}(c), which is in contrast to the result for $\gamma_{\ph}=0$ in Fig.~\ref{fig:SpWithoutDephasing}(c).
Here, in Fig.~\ref{fig:SpWithDephasing}(d), we can see that $S_{21}(\nu,\ds)$ has a main peak at $\nu=\omega_{21}$ in a similar manner to Fig.~\ref{fig:SpWithoutDephasing}(d).
This is reasonable because $S_{21}(\nu,\ds)$ is the spectrum directly emitted from the TLS.
On the other hand, in Fig.~\ref{fig:SpWithDephasing}(e), an emission peak appears at $\nu=\omega_{\c}$ in addition to the peak at $\nu=\omega_{21}$, in contrast to Fig.~\ref{fig:SpWithoutDephasing}(e).
The emission peak at $\nu=\omega_{\c}$ is a consequence of the AZE, as mentioned above.
Then, by the destructive interference shown in Fig.~\ref{fig:SpWithDephasing}(f), the two emission peaks at $\nu=\omega_{21}$ are canceled out almost completely.
As a result, the emission peak at $\nu=\omega_{\c}$ is highlighted in the total spectra [Fig.~\ref{fig:SpWithDephasing}(c)].
Here, the almost complete canceling of the emission peaks indicates that the interference is not spoiled by the pure dephasing.

Hence, the mechanism can be schematically illustrated in Fig.~\ref{fig:SpWithDephasing}(o), where the AZE dissipation channel is additionally opened by the pure dephasing, in comparison with Fig.~\ref{fig:SpWithoutDephasing}(o).
This AZE channel provides an alternative route of radiation at the frequency of $\omega_{\c}$.
Its relative influence is maximized especially when the two dissipation channels are canceled out by the destructive interference.
As a result, at the detuning of $-3.16~\meV$, the emission line at the TLS transition energy is strongly reduced with a simultaneous intensity enhancement at the cavity resonance.
This understanding is consistent with our numerical results that such a phenomenon can be still observed, even if $\gamma_{\ph}$ is much smaller than $(\gamma+\kappa)/2$ (Fig.~\ref{fig:3ueV} in Appendix~\ref{app:3ueV}).
Here, in this scenario, we should notice that the quality of interference for the Fano effect is {\em not} lost by the pure dephasing.
This is because the two dissipation channels are subjected to identical phase fluctuations by the pure dephasing, and therefore, the phase difference between the two channels is not affected.
At this stage, we can further notice that the insufficient suppression of the transition rate [Fig.~\ref{fig:SpWithDephasing}(b)] is not due to the loss of the quality of interference but in fact due to the appearance of the AZE dissipation channel.
We can thus conclude that the interference effect is itself tolerant to the pure dephasing though the transition rate, $W$, is smeared out by the pure dephasing.

In contrast, such an effect rapidly diminishes by changing the value of detuning.
For example, at $\omega_{21}-\omega_{\c}=-1.00~\meV$, the emission peak at $\nu=\omega_{21}$ again becomes bright, as shown in Fig.~\ref{fig:SpWithDephasing}(g).
This is because the interference requires comparable strength of the two relevant field amplitudes.
In fact, for the detuning of $-3.16~\meV$, we can find that $S_{21}(\nu,\ds)$ is well matched to $S_{\c}(\nu,\ds)$ in strength at $\nu=\omega_{21}$, as seen in Figs.~\ref{fig:SpWithDephasing}(d) and \ref{fig:SpWithDephasing}(e). 
However, this balance is lost by making $\omega_{\c}$ closer to $\omega_{21}$ because the amplitude via the cavity is enhanced in ratio by the resonance effect, as seen in Figs.~\ref{fig:SpWithDephasing}(h) and \ref{fig:SpWithDephasing}(i).
In consequence, the destructive interference becomes imperfect [Figs.~\ref{fig:SpWithDephasing}(j) and \ref{fig:SpWithDephasing}(g)].
At zero detuning, then, the difference caused by the pure dephasing becomes small in this calculation [Figs.~\ref{fig:SpWithoutDephasing}(k)--(n) and \ref{fig:SpWithDephasing}(k)--(n)].
For positive detuning ($\omega_{21}>\omega_{\c}$), finally, the impact of the AZE channel is suppressed by the constructive interference, in comparison with negative detuning ($\omega_{21}<\omega_{\c}$).

We have thus clarified the Fano effect on the emission spectra, especially with and without the pure dephasing.
Based on the above discussion, however, we finally make three remarks.
First, the strong reduction of the spectral intensity can also be expected by an appropriate amount of a non-radiative dissipation of the TLS, instead of the pure dephasing.
In this case, the non-radiative dissipation channel plays an alternative role for the AZE channel and the restriction of $\int\dd\nu S(\nu,\ds) = 1$ is eliminated.
Furthermore, the non-radiative dissipation does not disturb the interference for the Fano effect.
As a result, the destructive interference can highlight the influence of the non-radiative dissipation and the spectral intensity can be reduced strongly without the simultaneous intensity enhancement at the cavity resonance.
Second, we expect that the interference of the two channels is also tolerant to the pure dephasing of the cavity, i.e., the fluctuations of the cavity resonance, $\omega_{c}$.
This may be somewhat paradoxical at first glance by considering the mechanism for the robustness against the pure dephasing of the TLS.
However, the relevant field escaped through the cavity is driven at the frequency of $\omega_{21}$, mediated via the virtual photon excitation.
This can be interpreted as the classical forced oscillation, as explained above.
Hence, the phase of this field is less sensitive to the fluctuations of the cavity resonance under the detuned condition.
Finally, we expect that experimental demonstration of the Fano effect is possible within current cavity QED setups.
However, it would be still challenging to achieve the high degree of the overlap parameter, $\eta$.

\section{Conclusions and outlook}\label{sec:Conclusion}
We have presented a detailed analysis on the Fano effect in the cavity QED system, where the TLS is simply coupled with the single mode cavity.
Although such a system has been discussed by many authors in the past, the Fano effect has been implicitly neglected in most cases.
As a result, little was known about the Fano effect in this system.
In our view, one reason is the absence of a flexible and systematic approach, based on the modern theories of open quantum systems.
Therefore, in the early part of the present paper, we have first formulated the Makovian QME, based on the typical interaction Hamiltonians of the cavity QED system.
It was then shown that the cross terms of the individual interaction Hamiltonians yield a simple but unfamiliar type of Liouville superoperator.
Based on this treatment, we have found that the Fano effect can be successfully described and the Fano formula can be generalized over the weak and strong coupling regimes with the pure dephasing effect.
We have thus clearly shown that the Markovian QME approach is advantageous for the description of the Fano effect.

In the later part, on the other hand, we have focused on the emission spectra.
Based on the same interaction Hamiltonians, we have formulated the emission spectra, starting from the discussion on the simple intensity detection.
As a result, the emission spectra were expressed in a consistent manner with the QME.
Furthermore, it was numerically shown that the emission line at the TLS transition energy undergoes a strong intensity reduction by the destructive interference in collaboration with the pure dephasing effect.
This phenomenon can be observed even if the suppression of the transition rate is largely washed out by the pure dephasing.
By studying the decomposed spectra, then, we have clarified the underlying mechanism, and finally concluded that the interference between the two dissipation channels is itself tolerant to the pure dephasing, in contrast to the expectation that the impact of interference is sensitive to decoherence processes.
This is because the two dissipation channels are subjected to identical phase fluctuations, and therefore, the phase difference between the two channels is not affected.
The insufficient suppression of the transition rate can be attributed to the appearance of the AZE dissipation channel.

The results described in this paper would provide a fundamental and prototypical methodology to treat the Fano effect in various contexts, although the present study was devoted to the problem of the initially excited TLS in the cavity QED system.
One direction for future research is the lasing action by including the effect of excitation because the interference between the two channels is robust against the dephasing process.
Another interesting direction is a consideration of multiple emitters inside the cavity.
Superradiance and/or subradiance may be affected by the Fano effect.
It would be also interesting to study other relevant systems, such as circuit QED systems, optomechanical systems, and plasmon systems.
We believe that our approach can provide new insights on the Fano effect in a wide range of fields.

\begin{acknowledgments}
M.Y. greatly appreciates fruitful discussion with Prof. Susumu Noda and Dr. Takashi Asano at Kyoto University, Japan, where part of this work was done during the author's doctoral studies (2007--2010).
A.L. thanks Dr. Benjamin Dwir and Prof. Eli Kapon for valuable discussion of the observed phenomenon at EPFL, Switzerland. This work was supported by JSPS KAKENHI Grant No.~JP18K03454.
\end{acknowledgments}

\appendix
%======================================================
%   Derivation of BM-QME
%======================================================
\section{Derivation of Eq.~\eqref{eq:BM-QME}}\label{app:BM-QME}
Here, we derive Eq.~\eqref{eq:BM-QME}.
In the interaction picture with respect to $\oH_0$, the von Neumann equation for the total density operator $\vrho(t)$ can be written as 
%----------
% Equation
%----------
\begin{align}
\frac{\dd}{\dd t}\vrho(t) = -\ii[\vV_{\S}(t)+\vHSB(t),\vrho(t)].
\label{eq:vonNeumann}
\end{align}
%----------
Therefore, by inserting its integral form
%----------
% Equation
%----------
\begin{align}
\vrho(t) = \vrho(0) -\ii\int^{t}_{0} \dd s [\vV_{\S}(s)+\vHSB(s),\vrho(s)],
\label{eq:IntForm}
\end{align}
%----------
into Eq.~\eqref{eq:vonNeumann}, we obtain
%----------
% Equation
%----------
\begin{align*}
&\frac{\dd}{\dd t}\vrho(t)
    = -\ii[\vV_{\S}(t)+\vHSB(t),\vrho(0)]\nonumber\\
    &  -\int^{t}_{0} \dd s \left\{ [\vV_{\S}(t), [\vV_{\S}(s),\vrho(s)]] + [\vV_{\S}(t), [\vHSB(s),\vrho(s)]] \right\}\\
    &  -\int^{t}_{0} \dd s \left\{ [\vHSB(t), [\vV_{\S}(s),\vrho(s)]] + [\vHSB(t), [\vHSB(s),\vrho(s)]] \right\}.
\end{align*}
%----------
By applying the Born approximation, $\vrho(t) \simeq \vrho_{\S}(t)\otimes\orho_{\B}$, with $\Tr_{\B}[\oHSB \orho_{\B}]=0$, the trace over the bath eliminates the terms with an odd number of $\vHSB$; 
%----------
% Equation
%----------
\begin{align}
\frac{\dd}{\dd t}\vrho_{\S}(t)
    &= -\ii[\vV_{\S}(t),\vrho_{\S}(0)]
      -\int^{t}_{0} \dd s [\vV_{\S}(t), [\vV_{\S}(s),\vrho_{\S}(s)]] \nonumber\\
    &  -\int^{t}_{0} \dd s \Tr_{\B}[\vHSB(t), [\vHSB(s),\vrho_{\S}(s)\otimes\orho_{\B}]].
    \label{eq:appA:temp}
\end{align}
%----------
% This equation is true up to second order in $\oHSB$.
Now, from Eq.~\eqref{eq:IntForm}, we have
%----------
% Equation
%----------
\begin{align*}
\vrho_{\S}(t) = \vrho_{\S}(0) -\ii\int^{t}_{0} \dd s [\vV_{\S}(s),\vrho_{\S}(s)],
\end{align*}
%----------
and therefore, Eq.~\eqref{eq:appA:temp} yields
%----------
% Equation
%----------
\begin{align*}
&\frac{\dd}{\dd t}\vrho_{\S}(t)
    = -\ii[\vV_{\S}(t),\vrho_{\S}(t)]\nonumber\\
    &\quad -\int^{t}_{0} \dd \tau \Tr_{\B}[\vHSB(t), [\vHSB(t-\tau),\vrho_{\S}(t-\tau)\otimes\orho_{\B}]],
\end{align*}
%----------
where the integration variable is changed to $\tau = t - s$ in the second term.
By applying the Markovian approximation, the upper limit of the integral goes to infinity with $\vrho_{\S}(t-\tau) \simeq \vrho_{\S}(t)$.
We can thus obtain Eq.~\eqref{eq:BM-QME}.
In the degree of accuracy, this approach is true up to second order in $\oV_{\S}+\oHSB$, instead of $\oHSB$ alone.
This means that the higher order terms neglected in Eq.~\eqref{eq:BM-QME} still have terms in second order of $\oHSB$.
In this sense, this approach is different from the standard one~\cite{Breuer02}.
However, Eq.~\eqref{eq:BM-QME} is advantageous because the system operators of $\oHSB$ directly appear in the final form of Liouville superoperators; there is no need to decompose the system operators into the eigenoperators of $\oHS$.

%======================================================
%   The dissipator due to the Fano interference
%======================================================
\section{The dissipator due to the Fano interference}\label{app:FanoDissipator}
Here, we derive the dissipator [Eq.~\eqref{eq:FanoDissipator}] due to the Fano interference.
By substituting Eq.~\eqref{eq:DoubleCommutation} into Eq.~\eqref{eq:BM-QME}, the cross terms give
%----------
% Equation
%----------
\begin{align}
&-\int^{\infty}_{0}\dd\tau \Tr_{\B} [\vHSB^{(1)}(t) ,[\vHSB^{(2)}(t-\tau), \vrho_{\S}(t) \otimes \orho_{\B}]]\nonumber\\
&\quad = \frac{\gamma_{\F}(\omega_{\c})}{2}e^{-\ii\omega_{\c,21}t}(\oac\vrho_{\S}(t)\osgm_{+} - \osgm_{+}\oac\vrho_{\S}(t)) + \mathrm{h.c.},
\label{eq:AppB-1}
\end{align}
%----------
and
%----------
% Equation
%----------
\begin{align}
&-\int^{\infty}_{0}\dd\tau \Tr_{\B} [\vHSB^{(2)}(t) ,[\vHSB^{(1)}(t-\tau), \vrho_{\S}(t) \otimes \orho_{\B}]]\nonumber\\
&\quad = \frac{\gamma_{\F}(\omega_{21})}{2}e^{-\ii\omega_{\c,21}t}(\oac\vrho_{\S}(t)\osgm_{+} - \vrho_{\S}(t)\osgm_{+}\oac) + \mathrm{h.c.},
\label{eq:AppB-2}
\end{align}
%----------
where $\gamma_{\F}(\omega)$ is 
%----------
% Equation
%----------
\begin{align}
\gamma_{\F}(\omega) &\equiv 2\sum_{\ell}\xi_{\ell}\zeta^{*}_{\ell} \int^{\infty}_{0}\dd\tau e^{\ii(\omega - \omega_{\ell}\tau)}, \nonumber\\
&\simeq 2\pi\sum_{\ell}\xi_{\ell}\zeta^{*}_{\ell}\delta(\omega-\omega_{\ell}).
\label{eq:AppB-gammaF}
\end{align}
%----------
In the second line of Eq.~\eqref{eq:AppB-gammaF}, for simplicity, we have neglected the principal value contribution in the formula,
%----------
% Equation
%----------
\begin{align*}
\int^{\infty}_{0}\dd \tau e^{\ii(\omega - \omega_{\ell})} = \pi \delta(\omega - \omega_{\ell}) + \ii\mathcal{P}\frac{1}{\omega - \omega_{\ell}}.
% \simeq \pi \delta(\omega - \omega_{\ell}).
\end{align*}
%----------
Here, $\gamma_{\F}(\omega)$ is zero if the spatial radiation patterns between the TLS and the cavity mode are orthogonal because $\xi_{\ell}\zeta^{*}_{\ell}=0$. 
In contrast, $\gamma_{\F}(\omega)$ plays an important role when the two radiation patterns are close with each other.
Hence, in addition to the second assumption in Sec.~\ref{sec:Setup}, i.e., $\xi_{\ell} \simeq \bar{\xi}(\omega_{\ell})$ and $\zeta_{\ell} \simeq \bar{\zeta}(\omega_{\ell})$, we further introduce a phenomenological parameter $\eta$ ($0 \le \eta \le 1$) that describes the degree of the overlap between the two radiation patterns.
Eq.~\eqref{eq:AppB-gammaF} is then given by
%----------
% Equation
%----------
\begin{align}
&\gamma_{\F}(\omega) \simeq 2\pi e^{\ii(\theta_{21} -\theta_{\c} )} \sqrt{\eta} \sum_{\ell} |\bar{\xi}(\omega_{\ell})| |\bar{\zeta}(\omega_{\ell})| \delta(\omega-\omega_{\ell}) \nonumber\\
&\quad = 2\pi e^{\ii(\theta_{21} -\theta_{\c} )} \sqrt{\eta} \sum_{\ell} \sqrt{\frac{\gamma}{2\pi D(\omega)}}
 \sqrt{\frac{\kappa}{2\pi D(\omega)}} \delta(\omega-\omega_{\ell}) \nonumber\\
&\quad = e^{\ii(\theta_{21} -\theta_{\c} )} \sqrt{\eta\gamma\kappa},
\label{eq:AppB-4}
\end{align}
%----------
where Eq.~\eqref{eq:gamma and kappa} has been used.
As a result of Eqs.~\eqref{eq:AppB-1}, \eqref{eq:AppB-2}, and \eqref{eq:AppB-4}, we can obtain Eq.~\eqref{eq:FanoDissipator}.

%----------
% Figure
%----------
\begin{figure}
\centering
\includegraphics[width=.50\textwidth]{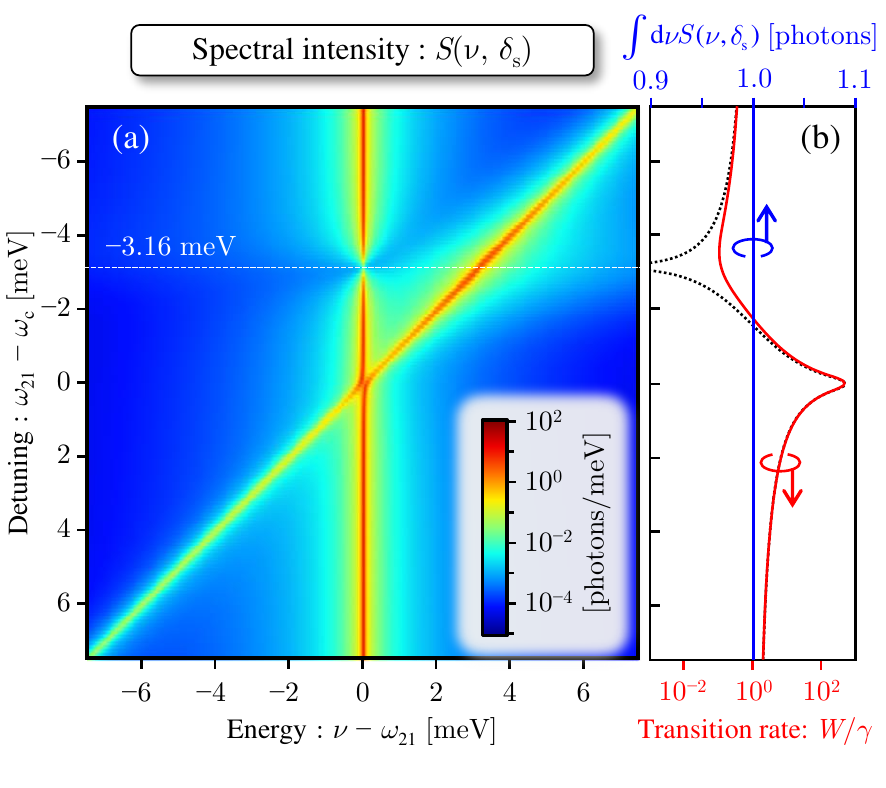}
\caption{
    Spectra for $\gamma_{\ph} = 3~\ueV$.
    Other parameters are the same as in Figs.~\ref{fig:SpWithoutDephasing} and \ref{fig:SpWithDephasing}.
    (a) The dependence of the spectral intensity, $S(\nu,\ds)$, on the detuning.
    (b) The transition rate, $W/\gamma$, and the integrated value, $\int \dd\nu S(\nu,\ds)$ [Eq.~\eqref{eq:nu-integration}], as a function of the detuning.
    For comparison, the transition rate for $\gamma_{\ph} = 0.0~\ueV$ is indicated by the dotted line, which is identical to the solid line in Fig.~\ref{fig:W}(a).
}
\label{fig:3ueV}
\end{figure}
%----------

%======================================================
%   The dissipator due to the Fano interference
%======================================================
\section{Numerical results for $\gamma_{\ph}=3~\ueV$}\label{app:3ueV}
We here show numerical results for $\gamma_{\ph}=3~\ueV$ in Fig.~\ref{fig:3ueV}.
In comparison with Fig.~\ref{fig:SpWithoutDephasing}(a), the spectra [Fig.~\ref{fig:3ueV}(a)] are largely changed although the modification of $W$ by the pure dephasing is weak ($\gamma_{\ph} \ll (\gamma+\kappa)/2$) [Fig.~\ref{fig:3ueV}(b)].
In this situation, the amount of pure dephasing is so small that the cavity feeding effect is limited, as can be seen for positive detuning ($\omega_{21}>\omega_{\c}$).
In contrast, we can find a strong intensity reduction at $\nu=\omega_{21}$ with a simultaneous intensity enhancement at the cavity resonance ($\nu=\omega_{\c}$) when $\omega_{21}-\omega_{\c} \sim -3.16~\meV$.
This is essentially the same phenomenon as we have seen in Fig.~\ref{fig:SpWithDephasing}(a).
Hence, we can notice that the key point for this observation is not the magnitude of the pure dephasing but the presence of the pure dephasing.
This is consistent with our scenario that the AZE channel additionally opened by the pure dephasing is highlighted by the destructive interference of the two dissipation channels.
This result also supports our understanding that the quality of interference for the Fano effect is {\em not} lost by the pure dephasing, as explained in the main text.

%======================================================
%   References and Notes
%======================================================
%--Note1--
\footnotetext{
    In the time evolution, $\dd p/\dd t$ oscillates around zero.
    Therefore, if we integrate both sides of Eq.~\eqref{eq:p} over a sufficient time scale, $\int (\dd p/\dd t) \dd t$ yields zero.
    After this treatment, we again differentiate both sides of the integrated equation.
    Then, the right hand side recovers that of Eq.~\eqref{eq:p}.
    However, the left hand side becomes zero.
    After this approximation, the time scale is coarse-grained by the integrated time scale.
    This time scale is roughly determined by the period of the Rabi oscillations when the system is in the strong coupling regime.
}

\newpage
% \bibliography{reference}
\input{Fano_ref.bbl}

\end{document}